\documentclass[aps,pra,twocolumn,superscriptaddress,longbibliography,footinbib]{revtex4-2}  %
\usepackage{graphicx} % needed for figures
\usepackage{float} % locate figures, [H]
\usepackage{dcolumn} % needed for some tables
\usepackage{bm,color}
\usepackage{amsmath,amssymb,dsfont,amstext,amsfonts}
\usepackage{extarrows}
\usepackage[colorlinks=true,linkcolor=blue,urlcolor=blue,citecolor=blue]{hyperref}
\usepackage{xcolor}
\usepackage{wasysym}
\usepackage{mathtools}
\usepackage{bbold}
\usepackage[export]{adjustbox}
\usepackage{mathdots}
\usepackage[normalem]{ulem} %makes underlining possible: \uline \uwave \sout
\usepackage{color}%makes \textcolor{grey}{text} available

\def\bra#1{\mathinner{\langle{#1}|}}
\def\ket#1{\mathinner{|{#1}\rangle}}

\def\braket#1{\mathinner{\langle{#1}\rangle}}

\def\sgn{\mathrm{sgn}}
\def\re{\mathrm{Re}\,}
\def\im{\mathrm{Im}\,}
\def\tr{\mathrm{Tr}}

\def\floor#1{\lfloor{#1}\rfloor}

\newcommand{\sect}[1]{\vspace{0.3em}{\it \textcolor{blue}{#1.---}}}
\begin{document}
\title{Localization via Quasi-Periodic Bulk-Bulk Correspondence}
\author{Dan S. Borgnia}
\email{dbognia@g.harvard.edu}
\affiliation{Department of Physics, Harvard University, Cambridge, MA 02138}
\author{Robert-Jan Slager}
\email{rjs269@cam.ac.uk}
\affiliation{TCM Group, Cavendish Laboratory, University of Cambridge, J. J. Thomson Avenue, Cambridge CB3 0HE, United Kingdom}
\affiliation{Department of Physics, Harvard University, Cambridge, MA 02138}
\date{\today}

\begin{abstract}
We report on a direct connection between quasi-periodic topology and the Almost Mathieu (Andre-Aubry) metal insulator transition (MIT). By constructing quasi-periodic transfer matrix equations from the limit of rational approximate projected Green's functions, we relate results from $\text{SL}(2,\mathbb{R})$ co-cycle theory (transfer matrix eigenvalue scaling) to consequences of rational band theory. This reduction links the eigenfunction localization of the MIT to the chiral edge modes of the Hofstadter Hamiltonian, implying the localized phase roots in a topological ``bulk-bulk" correspondence, a bulk-boundary correspondence between the 1D AAH system (boundary) and its 2D parent Hamiltonian (bulk). This differentiates quasi-periodic localization from Anderson localization in disordered systems. Our results are widely applicable to systems beyond this paradigmatic model.

\end{abstract}
\maketitle
\sect{Introduction} Quasi-periodic systems have played and continue to play a prominent role in condensed matter physics \cite{prodan2015,kraus2012topological,jitomirskaya2019critical,PhysRevB.101.014205, tbg2, spirals, refealhusemblquasi, slonghiphasetrannonher,tbg1}. The Andre-Aubry-Harper (AAH) model, or the almost Mathieu Operator in the mathematics community, is the prototypical example of quasi-periodic systems. In contrast to periodic or randomly disordered systems, the AAH model hosts many exotic properties, including a metal insulator transition (MIT), a well established duality between the two phases \cite{aubry1980annals,frohlich1990localization,goncalves2021hidden}, and a connection to a 2D quantum Hall system on a lattice, nearest neighbor (n.n.) Hofstadter model. 

Here, we report a new relation between this 2D quasi-periodic topology and the famous metal insulator transition. Following the ideas in Ref. \cite{paper1} and constructing a ``rational approximate" transfer matrix equation (TME) sequence from projected Green's functions (pGF) of the \textit{rational approximates}, we construct an explicit map between the $\text{SL}(2,\mathbb{R})$ co-cycle theory \cite{jitomirskaya1999metal,avila2006reducibility,avila2006solving,avila2009ten,avila2017sharp,jitomirskaya2012analytic} describing the eigenfunction localization and these higher dimensional topological invariants. In particular, we show that the quasi-periodic localization is an example of 1D bulk localized states reflecting a virtual 2D bulk topological invariant, see Table~\ref{tablephase}. This exotic behavior does not occur in fully disordered systems as non-trivial Chern markers are incompatible with localization \cite{marcelli2020localization}, thereby both adding to the rapporteur of exotic quasi-periodic phenomena, such as the 1D MIT, higher dimensional topology, self-duality, as well as lending a deeper explanation to the differences between random and quasi-periodic disorder. 

We proceed by reviewing the AAH model, including the known phase diagram~\cite{jitomirskaya1999metal,jitomirskaya1998anderson,avila2017sharp}, the connection between the 1D AAH model and the 2D Hofstadter model, and accordingly a discussion of its topological classification \cite{bellissard1982quasiperiodic,bellissard1986gaplabeling,kraus2012topological,prodan2015,frohlich1990localization,connes1994quasiconformal}. We 
then outline the projected Green's function method for constructing 2D rational approximate TMEs from Ref.~\cite{paper1} and detail a procedure for projecting 2D TMEs back into 1D quasi-periodic TMEs. This links 2D chiral topological edge modes to a lack of 1D TME solutions and reproduces the full AAH phase diagram directly from the quasi-periodic bulk-boundary correspondence, a ``bulk-bulk" correspondence.

\begin{table}[ht]
    \centering
    \begin{tabular}{|c|c|c|c|c|}
     \hline
      &Horizontal & Vertical  & Phase & Chiral Projection\\
      \hline
    $V<t$ &Yes &  No  &  Metal & One\\
    \hline
    $V>t$ &No & Yes & Insulator & All   \\
    \hline
    $V = t$ &Yes/No & Yes/No & Transition & N/A \\
    \hline
    \end{tabular}
    \caption{For Diophantine $\alpha$, table lists AAH phases, the convergent 2D parent Hamiltonian unit cell, and number of chiral edge modes from 2D parent Hamiltonian surviving the 1D projection. Chiral edge modes isolate spectrum for $V>t$.}
    \label{tablephase}
\end{table}

\begin{figure*}[ht]
    \centering
    \includegraphics[scale=.197]{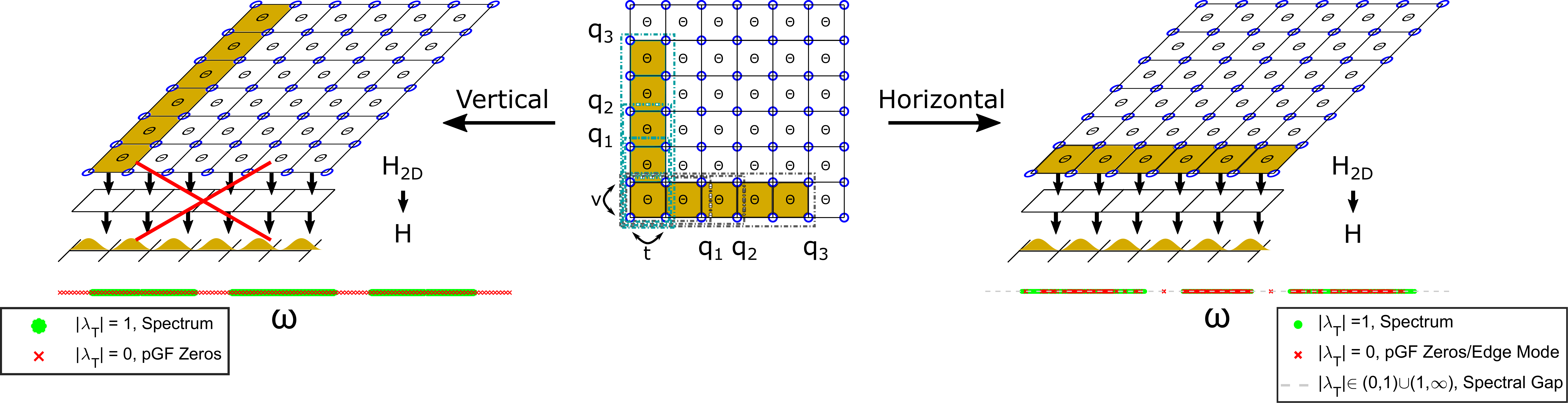}
    \caption{ Top panel shows Magnetic unit cells chosen in the 2D parent Hamiltonian (center). For Diophantine $\alpha$ in Eq. \eqref{2Dhamiltoniangauge}, the regimes $V<t$ and $V>t$ dictate a horizontal or vertical cell respectively, defining allowed projections back to 1D (left/right), i.e. horizontal unit cells naturally project back to 1D, while vertical unit cells do not. Bottom panel shows the transfer matrix eigenvalues corresponding to 1D projections and the according numerically computed AAH spectrum (green), corresponding spectral gaps (grey), and 2D pGF zeros (red x) for parameters $t =1, N = 2048, \alpha = \frac{1}{2}(\sqrt{5}-1)$ and $V = 2$ (left) or $V = 0.5$ (right). The spectrum (green) has transfer matrix eigenvalues on the unit circle, $\vert\lambda_{T}\vert=1$. Zero eigenvalues of the pGF (red x) imply the transfer matrix is rank deficient, $\lambda_T=0$. Note the many small gaps (grey) and zeros in almost (up to numerical precision) every gap for the horizontal projection, and the continuum of rank deficient points for the vertical projection.}
    \label{fig1}
\end{figure*}

\sect{Almost Mathieu Review} The AAH model presents a rich playground for new techniques in analysis and single-particle physics. Setting $\Theta = 2\pi \alpha$ and $\alpha\in\mathbb{R}-\mathbb{Q}$,
\begin{eqnarray}\label{HamiltonianAAHeq}
\hat{H} = \sum_{x} t(\hat{c}^{\dagger}_{x+1}\hat{c}_{x}+\hat{c}_{x+1}\hat{c}^{\dagger}_{x})+2V\cos(\Theta x+\delta)\hat{c}^{\dagger}_{x}\hat{c}_{x}.\quad
\end{eqnarray}

The model is a particular example of a 1D quasi-periodic system that has a non-trivial bulk topology, see S.I.~\ref{siaahalgebra} \cite{paper1,prodan2015}. We illustrate this with a 2D {\it parent} Hamiltonian formulation for the AAH model arising from phase shift degrees of freedom in the quasi-periodic pattern, i.e. $\delta \rightarrow\delta + \Theta$ is equivalent to $x\rightarrow x+1$ \cite{prodan2015,bellissard1982quasiperiodic,bellissard1986gaplabeling}. Stacking de-coupled AAH chains and parameterizing each layer by the phase choice $\delta_{y}$ \cite{jitomirskaya1998anderson,prodan2015,bellissard1982quasiperiodic,kraus2012topological} we can take an inverse Fourier transform along $\delta_{y}$. 
\begin{eqnarray}\label{2Dhamiltoniangauge}
    \mathcal{H}_{2D} &=&\sum_{x,\delta_{y}}t \hat{c}_{x+1,\delta_{y}}^{\dagger}\hat{c}_{x,\delta_{y}} + t^{*} \hat{c}_{x,\delta_{y}}^{\dagger}\hat{c}_{x+1,\delta_{y}}\nonumber\\
    &+& 2V\cos(\Theta x+\delta_{y})\hat{c}_{x,\delta_{y}}^{\dagger}\hat{c}_{x,\delta_{y}},\\
    \label{2Dhamiltonian}
    \tilde{\mathcal{H}}_{2D}  &=&\sum_{x,y}t (\hat{c}_{x+1,y}^{\dagger}\hat{c}_{x,y} + \hat{c}_{x,y}^{\dagger}\hat{c}_{x+1,y} )\nonumber\\
    &+& V(e^{i\Theta x}\hat{c}_{x,y+1}^{\dagger}\hat{c}_{x,y} + e^{-i\Theta x}\hat{c}_{x,y-1}^{\dagger}\hat{c}_{x,y}).
\end{eqnarray}
The result corresponds to a 2D tight-binding model with irrational magnetic flux per plaquette, $\Theta$, see Refs.~\cite{paper1,prodan2015}. Varying $\Theta$ produces the Hofstadter butterfly, Fig.~\ref{patternfigure}, with gaps labeled by integers, $\lbrace m+n\Theta\vert m,n\in\mathbb{Z}\rbrace$  \cite{bourne2018non}.

The AAH model famously hosts a 1D metal-insulator transition (MIT), first shown through the duality of Eq.~\eqref{HamiltonianAAHeq} under a Fourier transform-like, $\hat{c}_{k} = \sum_{x}\exp(i\Theta k x) \hat{c}_{x}$,
\begin{eqnarray}\label{HamiltonianAAHeqK}
\tilde{H} = \sum_{k} V(\hat{c}^{\dagger}_{k+1}\hat{c}_{k}+\hat{c}_{k+1}\hat{c}^{\dagger}_{k})+ 2t\cos(\Theta k+\delta_k) \hat{c}^{\dagger}_{k}\hat{c}_{k}.\quad 
\end{eqnarray}
The model is self-dual for $V = t$, fixing a transition from momentum-like (metallic) to position-like (insulating) eigenfunctions, S.I.~\ref{ogargumentsi} \cite{paper1,aubry1980annals}. 

A rigorous formulation of the MIT relies on the notion of \textit{spectral measure}, formally defined in S.I.~\ref{measureAAHsi} \cite{jitomirskaya1999metal}. Intuitively, a spectrum is absolutely continuous if the inner-product measure of an energy window $(\omega-\epsilon,\omega+\epsilon)$ continuously with $\epsilon$. This implies the presence of arbitrarily close eigenvalues which requires participation of arbitrarily many sites, guaranteeing de-localized eigenfunctions. By contrast, a spectrum is purely point-like if there is a discontinuity in the inner-product measure as a function of $\epsilon$. This implies localized eigenfunctions because discrete eigenvalues require fewer participating sites. For Almost-Mathieu operators, the spectrum is
\begin{enumerate}
    \item Absolutely continuous (metallic) for all $\Theta$ and all $\delta$ when $V < t$
    \item Singularly continuous (critical) for all $\Theta$ and all $\delta$ when $V = t$
    \item Pure point-like (insulator) for almost all $\Theta$ and almost all $\delta$ when $V > t$
\end{enumerate}
with the first ``almost" in (3) corresponding to Liouville irrationals (see below) and the second ``almost" corresponding to $\delta/2 = \Theta N+2\pi\mathbb{Z}$ for any $N\in\mathbb{N}$. 

These results were originally obtained through the construction of $\text{SL}(2,\mathbb{R})$ co-cycles -- transfer matrices indexed by scalars ($\delta$) -- associated with the AAH Hamiltonian, and developed results from co-cycle theory to classify the scaling behavior of transfer matrix solutions, i.e. Lyapunov exponents \cite{jitomirskaya1998anderson,jitomirskaya1999metal,jitomirskaya2012analytic,jitomirskaya2019critical,avila2006reducibility,avila2006solving,avila2009ten,avila2011holder,avila2017sharp}. 
Instead, we construct these same transfer matrices as the limit of a sequence of rational approximate transfer matrices following the methods recently developed in Ref.~\cite{paper1}

\sect{Rational Approximate Transfer Matrix Equations} We define a sequence of transfer matrix equations (TMEs) for quasi-periodic Hamiltonian operators whose quasi-periodic parameter, $\alpha\notin\mathbb{Q}$ is replaced by a rational $\frac{p_n}{q_n}\in\mathbb{Q}$ such that $\lim_{n\rightarrow\infty}\frac{p_n}{q_n} = \alpha$. The fastest converging sequence is the continued fraction approximation, 
\begin{align}\frac{p_n}{q_n} = a_0 + 1/\left(a_1 +  1/\left(a_2 + 1/\left( \ldots + 1/a_n \right) \right)\right).\end{align}
The golden mean is an example of a Diophantine number meaning its continued fraction approximation converges with at worst $\vert\alpha - \frac{p_n}{q_n}\vert<\frac{1}{\sqrt{5}q_n^{2}}$. By contrast, for Liouville numbers there exists a $\beta(\alpha)$ such that $\vert\alpha -\frac{p_n}{q_n}\vert < e^{-\beta(\alpha)q_n}$.

We will take sequences of rational approximate operators by substituting $\Theta = 2\pi\alpha$ in Eq.~\eqref{2Dhamiltoniangauge} with its continued fraction approximation $\Theta_N = 2\pi p_N/q_N$. For any rational flux, $\Theta/2\pi = p/q \in\mathbb{Q}$, one can define a magnetic unit cell specifying bands that have a Chern number, which sum to zero, and the irrational bulk topology is indeed the continued fraction limit.

For each rational approximate, $H_{2D,N}$, we construct a TME, see S.I.~\ref{transfermatrixsi} and- Refs. \cite{dwivedi2016bulk,paper1} for more details. The unit cell for the N-th rational approximate is $q_{N}$ sites long, and the eigenfunction on the n-th unit cell is 
\begin{eqnarray}
\Psi_{n,N} = \begin{pmatrix}\psi_{n+1}& \ldots & \psi_{n+q_{N}}\end{pmatrix}^{T}.
\end{eqnarray}
This translates the eigenvalue equation into
\begin{eqnarray}\label{egeq}
J_{N}\Psi_{n+1}+M_{N}\Psi_{n}+J_{N}^{\dagger}\Psi_{n-1} = E \Psi_{n}.
\end{eqnarray}
Here $J_{N}$ is the hopping matrix connecting the $q_{N}$-th site of $n$-th unit cell to the $1$-st site of the $n+1$-th unit cell, $J_{N}^{\dagger}$ does the opposite, and $M_{N}$ is the intra-unit cell term which acts internally on the $q_{N}$ internal sites of $\Psi_{n}$. 

For larger unit cells, the nearest neighbor (rank 1) hopping matrix, $J_N$, is not invertible. Taking advantage of the detailed work in \cite{dwivedi2016bulk}, we consider the reduced singular value decomposition of $J_N = V_{N}D_{N}W_{N}^{\dagger}$, and reduce Eq.~\ref{egeq} from a $2q_{N}\times 2q_{N}$ to a $2\times2$ matrix equation.

Defining, $G_{N} = (\omega - M_{N})^{-1}$ (projected Green's function) and $V_N,W_N$ by the reduced singular value decomposition, we construct our TME (setting $t =1$) \cite{dwivedi2016bulk, paper1},
\begin{widetext}
\begin{eqnarray}\label{tmeq}
\overbrace{(W_{N}^{\dagger}G_{N}V_{N})^{-1}\begin{pmatrix}
1 & -(W_{N}^{\dagger}G_{N}W_{N})\\
V_{N}^{\dagger}G_{N}V_{N} & V_{N}^{\dagger}G_{N}W_{N}(W_{N}^{\dagger}G_{N}V_{N})- V_{N}^{\dagger}G_{N}V_{N}W_{N}^{\dagger}G_{N}W_{N}
\end{pmatrix}}^{\hat{T}_{q_{N},n}}
\begin{pmatrix}
V_{N}^{\dagger}\Psi_{n} \\ W_{N}^{\dagger}\Psi_{n-1}
\end{pmatrix} = 
\begin{pmatrix}
V_{N}^{\dagger}\Psi_{n+1} \\ W_{N}^{\dagger}\Psi_{n}
\end{pmatrix}.
\end{eqnarray}
\end{widetext}
When $W_{N}^{\dagger}G_{N}W_{N}\neq 0$ and $V_{N}^{\dagger}G_{N}V_{N}\neq 0$, $\hat{T}_{q_N,n}$ is unitary and has reciprocal eigenvalues, $\lambda_{T,1}\lambda_{T,2} = 1$. The spectrum, $E\in\Sigma$, is formed by energies for which $\left\vert\lambda_T\right\vert = 1$. By contrast, energies for which $\left\vert\lambda_T\right\vert \in (0,1)\cup(1,\infty)$ form the spectral gaps, $E\in\mathbb{R}-\Sigma$, see Fig.~\ref{fig1} (bottom). 

The TMEs are determined by the translation invariant intra-cell projected Green's function (pGF), $G_{N}(\omega)$ in Eq.~\eqref{tmeq}. We construct it by inverse Fourier transforming the momentum-space bulk Green's function back into real space along one momentum dimension, $G(\omega,k_{\parallel}, x_{i},x_{f})$, acting on a single unit cell $x_{i} = x_{f}$,
\begin{align}\label{pgfeq}
\hat{G}_{\perp}(\omega,k_{\parallel},n)=\int dk_{\perp} G(\omega,k_{\parallel},k_{\perp})e^{ik_{\perp}\cdot (x_{f}-x_{i})},
\end{align}
In general, $k_{\parallel}, k_{\perp}$ represent momenta in parallel and transverse directions to the projection. The projection direction is fixed in 1D, but depends on the magnetic unit cell for the 2D rational approximates. E.g. a horizontal unit cell is the choice $G_N(\omega, \delta_y) = \hat{G}_{\perp,N}(\omega, k_{\parallel})$ with $\delta_y$ defined in Eq.~\eqref{2Dhamiltoniangauge}. 

The key point is that zeros of the pGF arise from topological winding of the bulk Green's function in Eq.~\ref{pgfeq} \cite{volovik2003universe,Rhim2018,Slager2015,Wilsons,mong2011edge,Borgnia2020} and correspond to edge modes, see S.I.~\ref{zerospolessi} and Refs. \cite{Slager2015,Borgnia2020,Rhim2018,paper1}. In Eq.~\ref{tmeq}, zeros of the pGF correspond to $W_{N}^{\dagger}G_{N}W_{N}= 0$ (or $V_{N}^{\dagger}G_{N}V_{N}=0$), such that  $\hat{T}_{q_N,n}$ is no longer unitary and $\lambda_T = 0$ (no normalizeable solutions). The pGF, $G_{N}(\omega)$, thus determines the existence of solutions to the rational approximate TMEs. However, we must check that as $N\rightarrow\infty$, these rational approximates converge to the irrational TME.

When rational pGFs converge to the irrational limit as discussed below, the band topology of the rational approximates can generate rank deficient points in the irrational TME, and the 2D parent Hamiltonian topology generates obstructions for the 1D quasi-periodic spectrum, see Fig.~\ref{fig1}(bottom). In practice, convergence/divergence depends on the parameter ratio $V/t$ and the choice of magnetic unit cell. For Diophantine $\alpha$, we find a horizontal (vertical) magnetic unit cell converges for $V < t$ ($V>t$) and diverges for $V>t$ ($V<t$), see Table~\ref{tablephase} and Figs.~\ref{fig1},~\ref{fig:convergence}. We note that while the unit cell choice appears symmetric, the original 1D system aligns with a horizontal unit cell arrangement.

\sect{Convergence Criteria}  Following the argument in \cite{paper1}, we check convergence by evaluating the (operator norm) difference between the rational pGF, $G_{N}(\omega)$, and the full irrational, $G_{\alpha}(\omega)$, defined by substituting $2\pi\alpha$ for $2 \pi\frac{p_{N}}{q_{N}}$ in $G_{N}(\omega)$.  We evaluate this difference by bounding (from above and below) the difference between the rational and irrational bulk Green's functions on a single unit cell, which reduces to bounding $\vert\vert(\omega-\tilde{H}_{2D,N})\vert\vert$ (S.I.~\ref{siboundstransfer}). 

The operator norm obeys the triangle inequality,
\begin{eqnarray}\label{triangleeq}
\left\vert\vert\omega\vert- \vert\vert\tilde{H}_{2D,N }\vert\vert\right\vert\leq\vert\vert(\omega-\tilde{H}_{2D,N})\vert\vert \leq\vert\omega\vert+\vert\vert\tilde{H}_{2D,N}\vert\vert\quad \quad
\end{eqnarray}
We can then  use the lower bound to prove divergence for $V>t$ and the upper bound to prove convergence for $V<t$ (for horizontal unit cell). Convergence and divergence are rigorously shown in S.I.~\ref{siboundstransfer}, but the main idea rests on the fact that if all \textit{Gershgorin circles},
\begin{eqnarray}
    \mathcal{R}_{i} = \left\lbrace r\in\mathbb{R}: \vert r - \tilde{H}_{2D,N,ii} \vert < t \right\rbrace
\end{eqnarray}
where $t = \sum_{j} \vert \tilde{H}_{2D,N,ij}\vert = 2t$ and $\tilde{H}_{2D,N,ij}$ are matrix elements of $\tilde{H}_{2D,N}$ -- of an infinite matrix overlap, the eigenvalues are contained in their union \cite{shivakumar1987eigenvalues}. Then for $V<t$, $\vert 2V\cos{\Theta x}\vert < 2t$ implies all circles overlap. 

However, when $V>t$, we prove the determinant of $H_{N}$ diverges with $q_{N}\rightarrow\infty$, S.I.~\ref{siboundstransfer}. This coupled with the fact that all but one eigenvalue can diverge \cite{jitomirskaya2019critical} implies the lower bound in Eq.~\eqref{triangleeq} diverges. Similarly, choosing a vertical unit cell inverts these cases and $V>t$ ($V<t$) leads to a bound (divergence), S.I~\ref{siboundstransfer}. This generates the phase diagram for Diophantine irrationals.

We obtain a more general convergence criteria by considering Liouville $\alpha$ with $\delta_N = e^{-\beta q_{N}}$. Even as the circles diverge, $\vert\vert H_{2D,N} \delta_{N}q_N \vert\vert$ converges for $V\lesssim te^{\beta}$. Thus, for $t<V<te^{\beta}$, both the horizontal and vertical unit cells are convergent, reproducing results from Ref.~\cite{avila2017sharp} as shown in Fig.~\ref{fig:convergence}. Note, vertical unit cells also converge for $te^{-\beta}<V<t$, but projection back to 1D (discussed below) keeps the phase diagram unaltered, see S.I.~\ref{gaugetransformsi}. 

Another subtle exception occurs when $\delta_{y}/2 = \Theta+2\pi\mathbb{Z}$, for which the 2D rational approximates, $H_{2D,N}$ have a chiral symmetry (class AIII) and are topologically trivial in 2D (no edge modes). In fact, any resonance $\delta_{y}/2 - k\Theta = 2\pi\mathbb{Z}$ suppresses the 2D edge modes and causes the MIT to vanish, again corroborating results in Ref.~\cite{avila2017sharp}.

\begin{figure}[ht]
    \centering
    \includegraphics[scale = 0.27]{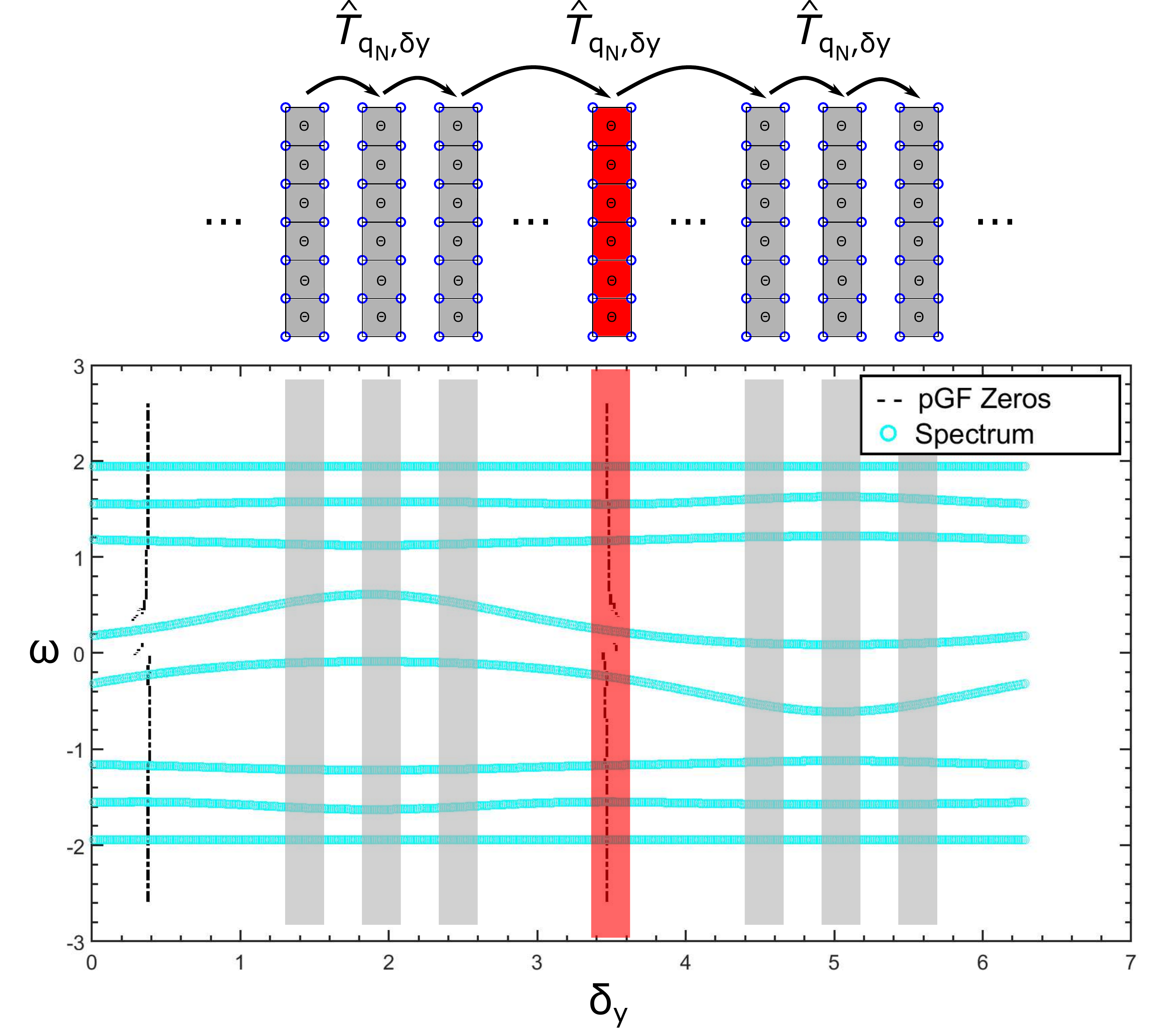}
    \caption{(top) Concatenated Transfer matrices for vertical unit cells. For each $\omega$ there exists a $\delta_y$ such that $\det{(\hat{T}_{q_{N},\delta_y})} =0$ (bottom) Rational approximate pGF zeros plotted as a function of $\delta_y$ ($k_y$) for $q_N = 8$ in golden mean approximation. Notice the presence of a zero for each energy (up to numerical precision) in the spectral gaps.}
    \label{fig:zeros}
\end{figure}

Aside from these subtle exceptions that in fact are invisible to numerical precision, bulk-boundary correspondence \cite{PhysRevLett.71.3697,prodan2015} guarantees the 2D rational approximates will host edge modes corresponding to the non-trivial magnetic flux per plaquette (the IQHE), see Fig.~\ref{fig:zeros}. Thus, any magnetic unit cell choice results in a rational pGF with zeros for each energy in the spectral gap, Fig~\ref{fig:zeros}. 

\sect{Return to 1D} Having generated a sequence of TME rational approximates in 2D, we construct the 1D physics by projecting back, fixing $\delta_{y}$. The 1D projection step notably is not symmetric between the two unit cell choices.

Projecting a horizontal unit cell into 1D is well defined for the rational approximates and in the irrational limit $N\rightarrow\infty$ (we are free to fix $\delta_{y}$ in the  Eq.~\eqref{2Dhamiltoniangauge}). Solutions to the horizontal unit cell TMEs thus form an absolutely continuous spectrum, the limit of the 2D spectrum. 

A vertical unit cell as constructed lacks a natural projection into 1D as the $\delta_y$ quantum number is unwrapped by the transfer matrix construction. To preserve $\delta_y$ in Eq.~\eqref{2Dhamiltoniangauge}, we choose vertical unit cells, but construct the rational TMEs by unwrapping the degenerate $k_x$ quantum number as done for horizontal unit cells, see Fig.~\ref{fig:zeros}(top). The corresponding rational pGF approximates still have a good $\delta_y$ quantum number, $G_{\perp,N}(\omega,\delta_y,x_n)$. However, the rational approximates do not grow in real-space and $\hat{T}_{q_N,n}$ spans a single real-space site, $x_n$, see Fig.~\ref{fig:zeros}. To avoid a dependence on $x_n$, we absorb the site information into the phase choice, $\delta_y(n) = \delta_y + \Theta n$ -- the same intuition as $\text{SL}(2,\mathbb{R})$ co-cycles \cite{avila2009ten,jitomirskaya1998anderson,jitomirskaya1999metal,avila2006reducibility}. We remark that doing otherwise will cause the resulting spectrum to deviate from the horizontal unit cell spectrum as guaranteed by the duality. Assuming $\delta_{y}/2 - k\Theta \neq 2\pi\mathbb{Z}$, all $\delta_y$ are included by the associated translations. We then concatenate the resulting transfer matrices into,
\begin{align}
    \hat{T}_{q_{N}} = \prod_{n =-\infty}^{\infty}\hat{T}_{q_N,n}(\delta_y(n)).
\end{align}
Notice, that $\det{(\hat{T}_{q_{N}})} = 0$ if for any $\delta_y$, $\det{(\hat{T}_{q_N,n})} = 0$. Thus if there exists an energy in the spectrum, $\omega \in \Sigma$, such that $\det{G_{\perp,N}(\omega,\delta_y)} = 0$ and $\det{(\hat{T}_{q_{N}})} = 0$, bulk normalizable solutions do not exist.

\sect{Topological Localization and emergent "bulk-bulk- correspondence} The 2D rational approximates are Hofstadter Hamiltonians with rational fluxes per plaquette hosting chiral edge modes. Correspondingly, the vertical rational approximate pGFs have zero eigenvalues for each $\omega$ in the rational spectral gaps, Fig.~\ref{fig:zeros}. This implies the rational transfer matrices have deficient points for every $\omega$ with at least one corresponding $\delta_y$. Since the rational pGF zeros converge in an operator norm sense to irrational pGF zeros, see S.I.~\ref{siboundstransfer}, the rational TME rank deficient energies correspond to irrational TME rank deficient energies. Thus, for a convergent vertical unit cell, the irrational 1D TME only has solutions at isolated energies with a discontinuity in transfer matrix rank across any spectral gap. This forms a pure-point like spectrum.

As discussed in S.I.~\ref{measureAAHsi} and Refs.~\cite{paper1,jitomirskaya1999metal,jitomirskaya1998anderson}, a pure point-like spectrum corresponds to localized eigenfunctions, while an absolute continuous spectrum corresponds to delocalized solutions. In this way, the transition from a horizontal to a vertical unit cell corresponds to a localization transition, and bulk-boundary correspondence -- the quasi-periodic bulk topology and the corresponding chiral edge modes -- forces bulk localization, a ``bulk-bulk" correspondence. In fact, the Fibonacci quasi-crystal lacks bulk-boundary correspondence \cite{prodan2015,bourne2018non} and, consistent with the discussion above, is power-law (not exponentially) localized for any parameter regime \cite{kalugin1986electron,capaz1990gap}. 

\sect{Conclusions} In summary, we report that the quasi-periodic metal insulator transition is indeed the consequence of bulk-boundary correspondence and quasi-periodic topology, generated by the chiral edge modes of the rational approximates. Beyond the novelty of topological bulk localization, the connection between quasi-periodic topology and eigenfunction localization makes clear the need to separate random disorder physics from quasi-periodic ``disorder" physics. While similar from the perspective of numerical simulation, random disorder does not impose these topological constraints.

Moreover, metallic eigenfunctions are invariant under phase shifts of the AAH on-site potential, while the localized eigenfunctions are not -- a phase shift is a translation and localized solutions are not translation invariant. This resembles a $U(1)$ symmetry breaking transition, as pointed out by Andre and Aubry in \cite{aubry1980annals,aubry1981bifurcation}.

In practice, this opens the door to many future questions regarding the role of topology in other applications of quasi-periodicity. The methods in this letter have already been shown to generalize beyond the AAH model \cite{paper1}, and we leave to future work the connection between localization and higher order topological invariants in higher dimensional quasi-periodic systems.

\begin{acknowledgements}
We cordially thank Ashvin Vishwanath for many helpful discussions and advice. We also thank Vir B. Bulchandani, Ruben Verresen, Matthew Gilbert, Nick G. Jones, Joaquin Rodriguez-Nieva, Dominic Else, Ioannis Petrides, Daniel E. Parker, Eitan Borgnia, Madeline McCann, William E. Conway, Saul K. Wilson, Will Vega-Brown, and especially Matthew Brennan for insightful discussions.
R.-J.S. acknowledges funding from the Winton Programme for the Physics of Sustainability and from the Marie Sk{\l}odowska-Curie programme under EC Grant No. 842901 as well as from Trinity College at the University of Cambridge.
\end{acknowledgements}
\bibliography{Refs}
%%% SUPPLEMENTAL %%% -----------------------------------------------------------------------------

\clearpage
\newpage
\pagebreak
\onecolumngrid
\begin{center}
\textbf{\large Supplemental Information - Localization via Quasi-Periodic Bulk-Bulk Correspondence}
\end{center}
\setcounter{equation}{0}
\setcounter{figure}{0}
\setcounter{table}{0}
\setcounter{page}{1}
\renewcommand{\theequation}{S\arabic{equation}}
\renewcommand{\thefigure}{S\arabic{figure}}

\begin{widetext}

\section{S.I. AAH Background}\label{measureAAHsi}
 The main results of this work generalize to multiple classes of quasiperiodic models, but the Andre-Aubry model (Almost-Mathieu operator) is the most well studied. We introduce it and some background on current methods in the study of single-particle quasiperiodic models. The Hamiltonian is simple, but presents a rich playground for new techniques in analysis and single-particle physics:
\begin{eqnarray}\label{HamiltonianAAHeqRS2}
\hat{H} = \sum_{x} t(\hat{c}^{\dagger}_{x+1}\hat{c}_{x}+\hat{c}_{x+1}\hat{c}^{\dagger}_{x})+2V\cos(\Theta x+\delta_{y})\hat{c}^{\dagger}_{x}\hat{c}_{x}.\quad
\end{eqnarray}
Here $\Theta = 2\pi a$, and we'll take $a\in\mathbb{R}-\mathbb{Q}$. 

\indent To understand why the Almost-Mathieu operator is mathematically interesting, beyond the physically interesting metal-insulator transition, we introduce the notion of a \textit{spectral measure}:

\noindent For any self-adjoint linear operator, $T$, one can decompose its measure on the target Hilbert space, $\mathcal{H}$ as an \textit{absolutely continuous}, \textit{singularly continuous}, and \textit{pure point-like} components. The spectral measure of $T$ is defined with respect to a vector $h\in\mathcal{H}$ and a positive linear functional $f: T\rightarrow \bra{h}f(T)\ket{h} = \int_{\sigma(T)}fd\mu_{h}$, where $\sigma(T)$ is the spectrum of the operator T and $\mu_{h}$ is the unique measure associated with $h$ and $T$.

The portion of the Hilbert space, i.e. the subspace of vectors, for which $\mu_{h}$ is dominated by the Lebesgue measure on the same subspace -- for every measureable set $A$, if the Lebesgue measure $L(A)= 0$, $\mu_{h}(A) = 0$ -- is absolutely continuous. By contrast, the pure point like component is the discrete portion of the spectrum where points can have finite measure in terms of $\mu_{h}$, but points have zero Lebesgue measure. The singularly continuous part of the spectrum is defined as the singular part of the spectrum -- the subspace which can be formed by a disjoint union of sets $A$ and $B$ for which $\mu_{h}(A) = 0$ when $L(B) = 0$ --  which is not pure point like.

\indent The original metal-insulator transition was shown non-rigorously through the duality of the Andre-Aubry under a Fourier transform-like operation, $\hat{c}_{k} = \sum_{x}\exp(i\Theta k x) \hat{c}_{x}$,
\begin{eqnarray}\label{HamiltonianAAHeqKS2}
{\tilde{\hat{H}}} = \sum_{k} V(\hat{c}^{\dagger}_{k+1}\hat{c}_{k}+\hat{c}_{k+1}\hat{c}^{\dagger}_{k})+ 2t\cos(\Theta k+\delta_{k}) \hat{c}^{\dagger}_{k}\hat{c}_{k}\quad,
\end{eqnarray}
see S.I.~\ref{ogargumentsi}. The model has a self-dual point for $V = t$, fixing a transition from momentum-like to position-like eigenfunctions. A more complete formulation of the problem, however, was constructed and proven for Almost-Mathieu operators. It was proven that the spectrum of the Almost-Mathieu operator is (setting $t = 1$)
\begin{enumerate}
    \item Absolutely continuous for all $\Theta$ and $\delta_{x}$ when $V < 1$
    \item Singularly continuous for all $\Theta$ and $\delta_{x}$ when $V = 1$
    \item Pure point-like for almost all $\Theta$ and $\delta_{x}$ when $V > 1$
\end{enumerate}
A pure point-like spectrum guarantees Anderson localization as it corresponds to eigenfunctions having finite measure at the eigenvalues and zero measure elsewhere. Intuitively, only a finite number can effectively participate (exponentially decaying weight) in generating discretely separated eigenvalues, and the eigenfunction is exponentially decaying on the lattice. More formally, the pure-point like spectrum forces eigenfunctions to be Semi-Uniformly-Localized-Eigenstates \cite{deift1983almost}. By contrast, an absolutely continuous spectrum guarantees delocalization if the spectrum has finite measure, which has been shown to be the case for the Almost-Mathieu operator. Much less is known about the singularly continuous case, and it has been the topic of multiple famous problems proposed by Barry Simon \cite{simon1984fifteen,simon2000schrodinger,simon2020twelve}. One of the few results on the singularly continuous spectrum is its existence deep in the pure-point like regime for Liouville $a = 2\pi/\Theta$ -- sequence of rational approximates $\lbrace p_n/q_n\rbrace$ exists such that $\vert a - p_{n}/q_{n}\vert < n^{-q_{n}}$ \cite{last2007exotic}. In fact, for Liouville numbers, the pure-point like transition occurs for $\lambda = e^{\beta}$ with $\beta = \lim_{n\rightarrow\infty}\ln(q_{n})/q_{n+1}$ \cite{avila2017sharp}.

In this language, the almost-Mathieu operator becomes a clear bridge between the well understood Mathieu operator (periodic operators) and random disorder. Understanding localization for the almost-Mathieu operator directly links to our understanding of chaos and localization in disordered systems. And yet, we still do not understand the full parameter space of a 1D nearest neighbor hopping lattice model with a cosine potential. The almost-everywhere part of this problem is important as it determines the physical stability of the model. Modern techniques in the field rely on cocycle theory  \cite{avila2006reducibility,avila2009ten,jitomirskaya1999metal,bourgain2002continuity}, and the absolutely continuous part of the spectrum is conjectured to be equivalent to the almost-reduciblity of the corresponding cocycle  \cite{avila2006reducibility}. The connection with cocycle theory further highlights the importance of this problem, as the reducibility classes of $SL(2,\mathbb{R})$ cocycles are known to describe the onset of quantum chaos and directly link to the Lyapunov exponent \cite{avila2006reducibility,avila2006solving,bourgain2002continuity}.

\section{S.I. The Andre Aubry's Argument}\label{ogargumentsi}
Early studies of quasi-periodic system dynamics focused on the construction of eigenstates from sequences of rational approximates, inductively \cite{dinaburg1975one,frohlich1990localization}. While the original work by Andre and Aubry \cite{aubry1980annals} relied on the continuity of the Thouless parameter and self-dual models to explain the transition. The RG like induction methods proved rigorously the existence of a localized phase. For the AAH model\cite{dinaburg1975one,frohlich1990localization} and similar quasiperiodic potentials\cite{frohlich1990localization}, these methods demonstrated the emergence of a \textit{pure point-like} spectrum for strong enough onsite potentials (relative to hopping terms). A pure point-like spectrum enforces localized eigenstates as eigenstates lack support across any continuous energy windows \cite{dinaburg1975one,frohlich1990localization,jitomirskaya1998anderson,jitomirskaya1999metal,jitomirskaya2019critical}. Below we introduce the simple AAH model and note key insights about the breakdown of Eigenstate ergodicity.

\indent The original paper by Andre and Aubry \cite{aubry1980annals} rests on two fundamental requirements for a quasi-periodic Hamiltonian, its self-duality and its fidelity to a sequence of rational approximates. It proposed a Hamiltonian, the AAH model, which satisfies a self-duality constraint under a real-space to dual-space (momentum space in the continuum limit) transformation, $\hat{c}_{k} = \sum_{x}\exp(i\Theta k x) \hat{c}_{x}$:
\begin{eqnarray}\label{aahrealspacetop}
	\hat{H} = \sum_{x} t(\hat{c}^{\dagger}_{x+1}\hat{c}_{x}+\hat{c}_{x+1}\hat{c}^{\dagger}_{y})+2V\cos(\Theta x+\delta_{x})\hat{c}^{\dagger}_{x}\hat{c}_{x}\quad\\
	{\tilde{\hat{H}}} = \sum_{k} V(\hat{c}^{\dagger}_{k+1}\hat{c}_{k}+\hat{c}_{k+1}\hat{c}^{\dagger}_{k})+2t\cos(\Theta k+\delta_{k})\hat{c}^{\dagger}_{k}\hat{c}_{k}\quad\label{aahkspacetop}
\end{eqnarray}
\indent Here $\Theta$ is some irrational parameter relative to $\pi$ and clearly for $t = V$ the Hamiltonian is self-dual, indicating the existence of a transition. One can introduce a sequence of rational approximates, $\lbrace a_{n}/b_{n}\rbrace_{n\in\mathbb{N}}$ with $a_{n},b_{n}\in\mathbb{Z}$ and $\lim_{n\rightarrow\infty} a_{n}/b_{n} = \Theta$. The sequence of Hamiltonians with periodic potentials links the density of states on either side of the duality transformation because there are well-defined bands. One can then write down the corresponding Thouless exponent for each side of the transition:
\begin{eqnarray}
\gamma(E) = \int_{-\infty}^{\infty}\log\vert E-E'\vert dN(E')\label{thouless_exp_gen}
\end{eqnarray}
For rational $a_{n}/b_{n}$, with $t = 1$ and $V = \lambda$, the transformation from Eq.~\eqref{aahrealspacetop} to Eq.~\eqref{aahkspacetop} takes $\tilde{V}\rightarrow1/\lambda$ and $E\rightarrow \tilde{E}/\lambda$, which implies $N_{\lambda,k} (E) = \tilde{N}_{1/\lambda,k}(E/\lambda)$ \cite{aubry1980annals}. So,
\begin{eqnarray}\label{thoulessduality}
\gamma(E) &=& \int_{-\infty}^{\infty}\log\vert \frac{E-E'}{\lambda}\vert d\tilde{N}\left(\frac{E'}{\lambda}\right)  +\log\vert\lambda\vert\nonumber\\
\gamma(E) &=& \tilde{\gamma}(\frac{E}{\lambda})+\log\vert\lambda\vert.
\end{eqnarray}
Since quasiperiodic systems do not have bands, but rather protected band gaps (discussed below \cite{jitomirskaya1998anderson,jitomirskaya1999metal,avila2011holder,zhao2020holder,amor2009holder,jitomirskaya2019critical}), the Thouless exponent must be non-negative by construction in 1D \cite{aubry1980annals}. Thus, for $\lambda>1$, $\gamma(E)>0$ and states are exponentially localized. This all relies on the continuity of the Thouless exponent, only proven in 2002 \cite{bourgain2002continuity}. Here, the rational approximates differ drastically from the irrational limit, but the density of state is well described by the approximation. In fact, these spectral properties are topologically protected by the quasiperiodic pattern's robustness \cite{prodan2015}, further expanded below.

\indent Via the above arguments, Andre and Aubry demonstrated the existence of non-zero Thouless parameter for $V >1$. And, by the duality of the model, $\gamma(E)$ must be zero for $V<1$. This transition is unusually sharp, exhibiting exponential localization on either side due to the relation between the Thouless parameters of the self-dual models. Further, the methodology is quite general in 1D and can be extended to other self-dual models, even if the duality is energy dependent, \cite{paper1}. The argument breaks down in higher dimensions as the Thouless exponent is no longer guaranteed to be non-negative \cite{aubry1980annals}. 

\indent Returning to the Hamiltonian in Eq.~\ref{aahrealspacetop}, note the phase $\delta_{x}$ sets the "origin" of the pattern. The eigenvalues cannot depend on the phase $\delta_{x}$ in the thermodynamic limit. However, for $\lambda >1$, if a state of energy $E$ is localized to site $x$ when $\delta_{x} = 0$, then the state localized to site $x-\delta/\Theta$ has energy $E$ for the shifted Hamiltonian with phase $\delta_{x} = \delta$. Thus, the eigenstates of each eigenvalue do depend directly on the phase. In \cite{aubry1980annals,aubry1981bifurcation}, this is described as a gauge-group symmetry breaking transition. \\

\section{S.I. Transfer Matrix}\label{transfermatrixsi}
The explicit construction of a $2\times2$ transfer matrix starts with a reduced SVD of the hopping matrix $J_{N}$,
\begin{eqnarray}
J_{N} &=& V_{N} D_{N} W^{\dagger}_{N}\\
J_{N}^{\dagger} &=& W_{N} D_{N}^{\dagger} V^{\dagger}_{N}
\end{eqnarray}
with $V^{\dagger}V = W^{\dagger}W = \mathbb{1}$ and $W^{\dagger} V = 0$, and in the case of $J_{N}$, 
\begin{eqnarray}
D_{N} = D_{N}^{\dagger}=\begin{pmatrix}
t & 0 & \ldots &0\\
0 & 0 & \ldots & 0\\
\vdots& \vdots&\ddots & \vdots\\
0 & 0 & \ldots & 0
\end{pmatrix},
\end{eqnarray}
Since our hopping matrix is rank 1, we truncate $D_{N} = t$ and and correspondingly, the $q_{N}\times 1$ dimensional operators
\begin{eqnarray}
W_{N}= \begin{pmatrix}
0 \\ \vdots \\ 0 \\(-1)^{q_{N}}
\end{pmatrix},
V_{N}= \begin{pmatrix}
(-1)^{q_{N}} \\0\\  \vdots \\ 0
\end{pmatrix}
\end{eqnarray}
Rewriting our unit cell Green's function as 
\begin{eqnarray}
G_{N}(\omega) = (\omega - M_{N})^{-1}
\end{eqnarray}
The transfer matrix equation reduces to
\begin{eqnarray}
\Psi_{n} &=& G_{N}J_{N}\Psi_{n+1} + G_{N}J_{N}^{\dagger}\Psi_{n-1}\\
\Psi_{n} &=& G_{N}V_{N}D_{N}W^{\dagger}_{N}\Psi_{n+1} +  G_{N}W_{N}D_{N}V^{\dagger}_{N}\Psi_{n-1}
\end{eqnarray}
Projecting into the $V_{N},W_{N}$ subspaces of $\Psi_{n}$,
\begin{eqnarray}
V^{\dagger}_{N}\Psi_{n} &=& V^{\dagger}G_{N}V_{N}D_{N}W^{\dagger}_{N}\Psi_{n+1} +  V^{\dagger}G_{N}W_{N}D_{N}V^{\dagger}_{N}\Psi_{n-1}\\
W^{\dagger}_{N}\Psi_{n} &=& W^{\dagger}G_{N}V_{N}D_{N}W^{\dagger}_{N}\Psi_{n+1} +  W^{\dagger}G_{N}W_{N}D_{N}V^{\dagger}_{N}\Psi_{n-1}.
\end{eqnarray}
This reduces the Transfer matrix equation to \cite{dwivedi2016bulk}, setting $t =1$,
\begin{align}
\begin{pmatrix}
(W_{N}^{\dagger}G_{N}V_{N})^{-1} & -(W_{N}^{\dagger}G_{N}V_{N})^{-1}(W_{N}^{\dagger}G_{N}W_{N})\\
(V_{N}^{\dagger}G_{N}V_{N})(W_{N}^{\dagger}G_{N}V_{N})^{-1} & V_{N}^{\dagger}G_{N}W_{N}- V_{N}^{\dagger}G_{N}V_{N}(W_{N}^{\dagger}G_{N}V_{N})^{-1}W_{N}^{\dagger}G_{N}W_{N}
\end{pmatrix}\begin{pmatrix}
V_{N}^{\dagger}\Psi_{n} \\ W_{N}^{\dagger}\Psi_{n-1}
\end{pmatrix} = 
\begin{pmatrix}
V_{N}^{\dagger}\Psi_{n+1} \\ W_{N}^{\dagger}\Psi_{n}
\end{pmatrix}
\end{align}
Notice all of the elements in the $2\times2$ transfer matrix are effectively scalars and thus commute with each other and we can just factor out the common factor $(W_{N}^{\dagger}G_{N}V_{N})^{-1}$,
\begin{eqnarray}
(W_{N}^{\dagger}G_{N}V_{N})^{-1}\begin{pmatrix}
1 & -(W_{N}^{\dagger}G_{N}W_{N})\\
V_{N}^{\dagger}G_{N}V_{N} & V_{N}^{\dagger}G_{N}W_{N}(W_{N}^{\dagger}G_{N}V_{N})- V_{N}^{\dagger}G_{N}V_{N}W_{N}^{\dagger}G_{N}W_{N}
\end{pmatrix}
\begin{pmatrix}
V_{N}^{\dagger}\Psi_{n} \\ W_{N}^{\dagger}\Psi_{n-1}
\end{pmatrix} = 
\begin{pmatrix}
V_{N}^{\dagger}\Psi_{n+1} \\ W_{N}^{\dagger}\Psi_{n}
\end{pmatrix}
\end{eqnarray}
Now, if ever $(V_{N}^{\dagger}G_{N}V_{N})=0$ and $(V_{N}^{\dagger}G^{2}_{N}V_{N})=0$ or $(W_{N}^{\dagger}G_{N}W_{N})=0$ and $(W_{N}^{\dagger}G_{N}^{2}W_{N})=0$ the transfer matrix has rank 1 and the only non-vanishing solution is localized at the edge. The corresponding energy is not in the bulk spectrum, $\omega\notin\Sigma$. The above condition happens when $G_{N}(\omega) = (\omega-M_{N})^{-1} = 0$. These zeros correspond to protected obstructions of the spectrum and are important to quasi-periodic localization.

\section{S.I. Bulk to Projected Green's Function}\label{siboundstransfer}
Our rational approximation to the irrational Hamiltonian follows the continued fraction approximation of irrational parameter $\alpha$ described above. 
We define,
\begin{eqnarray}
\hat{H}_{N} = \sum_{x}t(\hat{c}^{\dagger}_{x+1}\hat{c}_{x}+\hat{c}^{\dagger}_{x}\hat{c}_{x+1})+2V\cos{(2\pi \frac{p_{N}}{q_{N}}x+\phi)}\hat{c}_{x}^{\dagger}\hat{c}_{x}
\end{eqnarray}
for which we can define a $q_{N}$ site unit cell and Fourier transform into (setting $t =1$)
\begin{align}\label{fullmatrix}
\tilde{H}_{N} = \sum_{k}e^{ikx}c_{k}^{\dagger}c_{k}\begin{pmatrix}
2V\cos{(2\pi \frac{p_{N}}{q_{N}}1+\phi)} & 1 &0& \ldots & e^{ik}\\
1 & 2V\cos{(2\pi \frac{p_{N}}{q_{N}}2+\phi)} & 1& \ldots & 0\\
0 & 1 & 2V\cos{(2\pi \frac{p_{N}}{q_{N}}3+\phi)}& \ddots& \vdots\\
\vdots & \vdots &\ddots& \ddots  & 1\\
e^{-ik}  & \ldots  &0& 1 & V\cos{(2\pi \frac{p_{N}}{q_{N}}q_{N}+\phi)}
\end{pmatrix}.
\end{align}
The bulk Green's function is just the natural $G(\omega,k) = (\omega-\tilde{H}_{N}(k))^{-1}$ and the corresponding ``projected" Green's function is simply
\begin{eqnarray}
G_{\perp,N}(\omega)=\int \frac{dk}{2\pi}G(\omega,k)
\end{eqnarray}
We examine difference between the almost tri-diagonal matrix in E.Q.~\eqref{fullmatrix} and the irrational Hamiltonian on a given unit cell.
First, we set $\phi = 0$ w.l.o.g. and notice that 
\begin{eqnarray}
V\cos(2\pi\alpha x) = V\cos(2\pi(\frac{p_{N}}{q_{N}}+\delta_{N}) x) = 2V\cos(2\pi\frac{p_{N}}{q_{N}} x)- 2V\sin{(\pi\delta_{N}x)} \sin{(2\pi\frac{p_{N}}{q_{N}} x+\pi\delta_{N}x)}
\end{eqnarray}
So,
\begin{eqnarray}\label{diffbound}
\vert 2V\cos(2\pi\alpha x)-2V\cos(2\pi\frac{p_{N}}{q_{N}} x)\vert < \vert4V\sin{(\pi\delta_{N}x)}\vert = \vert 4V\sum_{n = 1}^{\infty} \frac{(-1)^{n}}{(2n-1)!} (\pi\delta_{N}x)^{2n-1}\vert<\vert 4V\pi\delta_{N}q_{N}\vert
\end{eqnarray}
where in the last inequality we used that an alternating and uniformly convergent series is bounded at any step and that $x\leq q_{N}$. Notice, that for any irrational number $\delta_{N}<\frac{1}{\sqrt{5}q_{N}^{2}}$ and thus
$$\vert 4V\pi\delta_{N}q_{N}\vert < \frac{4V\pi}{\sqrt{5}q_{N}}\rightarrow 0$$ in the limit of large $q_{N}$.
However, we keep $\delta_{N}$ going forward, to account for special cases of stronger bounds on $\delta_{N}$, i.e. Liouville numbers. Expanding the full irrational parameter on-site Green's function,
\begin{eqnarray}
G_{\perp,\alpha}(\omega)=\int\frac{dk}{2\pi}G_{N}(\omega)\left(\mathbb{1} + G_{N}(\omega)\left(\sum_{n=1}^{q_{N}}\left[2V\cos(2\pi\alpha n)-2V\cos(2\pi\frac{p_{N}}{q_{N}} n)\right]\ket{n}\bra{n}\right)\right)^{-1}
\end{eqnarray}
we see that
\begin{eqnarray}
G_{\perp,\alpha}(\omega)-G_{\perp,N}(\omega) &=& -\int\frac{dk}{2\pi}G_{N}^{2}(\omega)
\frac{\left(\sum_{n=1}^{q_{N}}\left[2V\cos(2\pi\alpha n)-2V\cos(2\pi\frac{p_{N}}{q_{N}} n)\right]\ket{n}\bra{n}\right)}{\left(\mathbb{1} + G_{N}(\omega)\left(\sum_{n=1}^{q_{N}}\left[2V\cos(2\pi\alpha n)-2V\cos(2\pi\frac{p_{N}}{q_{N}} n)\right]\ket{n}\bra{n}\right)\right)}.
\end{eqnarray}
We can bound the above expression in terms of the operator norm by using our above bound on
$$\vert 2V\cos(2\pi\alpha x)-2V\cos(2\pi\frac{p_{N}}{q_{N}} x)\vert<\vert4V\pi\delta_{N}q_{N}\vert.$$
Convergence then depends on the uniform convergence of 
\begin{eqnarray}
\left\Vert G_{N}(\omega)\left(\sum_{n=1}^{q_{N}}\left[2V\cos(2\pi\alpha n)-2V\cos(2\pi\frac{p_{N}}{q_{N}} n)\right]\ket{n}\bra{n}\right)\right\Vert^{k}<\Vert\left(G_{N}(\omega)4V\pi\delta_{N}q_{N}
\right)\Vert^{k}
\end{eqnarray}
as $k\rightarrow\infty$, where we have used the operator norm $\vert\vert A\vert\vert = \sup_{\psi\in \mathcal{H}} (\vert \vert A\psi\vert\vert/\vert\vert\psi\vert\vert)$. We need
\begin{eqnarray}
\Vert\left(G_{N}(\omega)4V\pi\delta_{N}q_{N}
\right)\Vert < 1
\end{eqnarray}
We have taken $\omega$ not in the spectrum, $\Sigma$ of the operator and $G_{N}(\omega)<\infty$, but this doesn't necessarily bound $G_{N}(\omega)<1/(\delta_N q_N)$. The operator norm of $G_{N}(\omega)$ is at most the inverse of the gap half-width. The minimum gap width for the rational Hofstadter approximates is known, when $V<1$, $\sim V^{q_N/2}$ \cite{jitomirskaya2020spectrum}, which means $G_{N}(\omega) \sim V^{-q_N/2}$. Taking this approach, there will always be gaps for which the pGF doesn't converge.

Instead, consider a small shift of $\omega$ towards the upper and lower half of complex plane $\omega\rightarrow\omega \pm i\epsilon$. This fixes the maximum of $G_{N}(\omega\pm i\epsilon) < \epsilon^{-1}$. We can always take a large enough $q_N$ such that $\delta_{N}q_{N} < \epsilon$ for any $\epsilon>0$. However, we now need both the imaginary and real parts of the offset pGF to converge.
\begin{align}
    \re{(G_{\perp,\alpha}(\omega\pm i\epsilon) - G_{\perp,N}(\omega\pm i\epsilon))} = -\int\frac{dk}{2\pi}\frac{\left[(\omega-\tilde{H}_{N})^{2} + (\omega-\tilde{H}_{N})(\tilde{H}_{N}-\tilde{H}_{\alpha})-\epsilon^{2}\right](\tilde{H}_{\alpha}-\tilde{H}_{N})}{((\omega-\tilde{H}_{N})^{2}+\epsilon^{2})^{2}\left(\mathbb{1}+\frac{2\omega (H_{N} - H_{\alpha})+\tilde{H}_{N}^{2}- \tilde{H}_{\alpha}^{2}}{((\omega-\tilde{H}_{N})^{2}+\epsilon^{2})}\right)}\\
    \im{(G_{\perp,\alpha}(\omega\pm i\epsilon) - G_{\perp,N}(\omega\pm i\epsilon))} = \int\frac{dk}{2\pi}\frac{\left[\pm 2i\epsilon(\omega-\tilde{H}_{N})+i\epsilon(H_{N} -\tilde{H}_{\alpha})\right](\tilde{H}_{\alpha}-\tilde{H}_{N})}{((\omega-\tilde{H}_{N})^{2}+\epsilon^{2})^{2}\left(\mathbb{1}+\frac{2\omega (H_{N} - H_{\alpha})+\tilde{H}_{N}^{2}- \tilde{H}_{\alpha}^{2}}{((\omega-\tilde{H}_{N})^{2}+\epsilon^{2})}\right)}
\end{align}
Now we use that $\vert\vert(\omega-\tilde{H}_{N})^{2}+\epsilon^{2}\vert\vert >\epsilon^{2}$, to cancel the bottom factors $\sim \frac{1}{\epsilon^{2}}$. And, we can use the bound from above, $\vert\vert\tilde{H}_{N} - \tilde{H}_{\alpha}\vert\vert<4V\pi\delta_N q_N$ to get 
\begin{align}
   \vert\vert \re{(G_{\perp,\alpha}(\omega\pm i\epsilon) - G_{\perp,N}(\omega\pm i\epsilon))}\vert\vert < \int\frac{dk}{2\pi}\vert\vert\epsilon^{-4}\left[-\epsilon^{2}4V\pi\delta_N q_N+ (\omega-\tilde{H}_{N})\left(4V\pi\delta_N q_N\right)^{2} \right]\vert\vert\\
    \vert\vert\im{(G_{\perp,\alpha}(\omega\pm i\epsilon) - G_{\perp,N}(\omega\pm i\epsilon))}\vert \vert< \int\frac{dk}{2\pi}\epsilon^{-3}\left[ 8 V\pi\delta_N q_N(\omega-\tilde{H}_{N})+\left(4V\pi\delta_N q_N\right)^{2}\right].
\end{align}
Clearly the imaginary part will converge to zero for large $q_N$. We bound the real part by showing
\begin{eqnarray}\label{ourboundeq}
\vert\vert(\omega -\tilde{H}_{N})\vert\vert < C q_N,
\end{eqnarray}
so that we can always take $q_N$ big enough to make $\left(4V\pi\delta_N q_N\right) C q_N<\epsilon^{3}$ for any $\epsilon>0$ to get the uniform convergence above, for all Diophantine $\alpha$.

The operator norm obeys the triangle inequality,
\begin{eqnarray}\label{triangleq}
\left\vert\vert\omega\vert- \vert\vert\tilde{H}_{N}\vert\vert\right\vert\leq\vert\vert(\omega-\tilde{H}_{N})\vert\vert \leq\vert\omega\vert+\vert\vert\tilde{H}_{N}\vert\vert
\end{eqnarray}
We can then  use the lower bound to prove divergence for $V>1$ and the upper bound to prove convergence for $V<1$. 

While any finite matrix has finite eigenvalues contained in Gershgorin circles,
$\mathcal{R}_{i} = \left\lbrace r\in\mathbb{R}: \vert r -\tilde{H}_{N,ii} \vert < t \right\rbrace$
where $t = \sum_{j} \vert \tilde{H}_{N,ij}\vert = 2t$, the limit of infinite matrices are not as simple. Indeed, in \cite{jitomirskaya2019critical} the semi-infinite spectrum of $\tilde{H}_{N}$ is noted to contain up to 2 eigenvalues inside each gap and up to 1 in each of the infinite intervals above and below the bulk. We need to bound the eigenvalues above and below the gap. We will bound them as constant when $V<t$ and as exponentially divergent for $V>t$

Fortunately, for $V<t$ all Gershgorin circles intersect for all $q_N$ as $\vert2\cos{\Theta x}\vert< 2t$. Thus, we are guaranteed that the infinite spectrum lies in the union of the circles. By contrast, for $V>t$ there exist Gershgorin circles which do not intersect, as $\max_x\vert2\cos{\Theta x}\vert > 2t$. Here we can even show a divergence of the largest eigenvalue, by considering the determinant of $\tilde{H}_N$. In particular, if $V>t$ and $t = 1$
\begin{align}
    \det\tilde{H}_{N} = \sum_{1\geq n \leq q_{N}} \binom{q_{N}}{n}(2V)^{n}\left(\int_{0}^{2\pi} \prod_{j =1}^{n}(dx_{j} \cos{x_j})\right) > \sum_{n \leq \floor{q_N/2} } \binom{q_{N}}{2n}V^{2n} > 2^{\floor{q_{N}/2}}
\end{align}
with odd terms canceling in the integral. As the determinant diverges, the determinant of $\tilde{H}_{N}^{-1}$ approaches 0 as $q_{N}\rightarrow\infty$, implying the existence of zero eigenvalues, implying $\vert\vert\tilde{H}_{N}\vert\vert$ is unbounded in the limit of $q_{N}\rightarrow\infty$.

Thus, for $V<1$ and any $\epsilon>0$ both the imaginary and real parts of the rational pGFs converge for $q_N$ such that $\delta_{N}q_{N}<\epsilon^{4}$. This implies the uniform convergence of the rational approximates and the Green's function of the irrational limit is arbitrarily close to that of the translation invariant Green's function. Using the transfer matrix construction above, the quasi-periodic transfer matrix is arbitrarily close to the translation invariant transfer matrix. This is similar to almost-reducibility as defined by Artur Avilla in \cite{avila2006reducibility}.

Now, consider the case of $\delta_{N} = e^{-\beta q_{N}}$, when $\alpha$ is Liouville. Here,
\begin{eqnarray}
\Vert\left(G_{N}(\omega)4V\pi\delta_{N}q_{N}
\right)\Vert < \Vert\left(G_{N}(\omega)4V\pi e^{-\beta q_{N}}q_{N}
\right)\Vert 
\end{eqnarray}
Now, even if $V>1$ we can bound the determinant of $\det\tilde{H}_{N} \delta_{N}q_{N}$
\begin{eqnarray}
    \det\tilde{H}_{N} \delta_{N}q_{N} = \sum_{1\geq n \leq q_{N}} \binom{q_{N}}{n}(2V)^{n}\left(\int_{0}^{2\pi} \prod_{j =1}^{n}(dx_{j} \cos{x_j})\right) < \sum_{n \leq \floor{q_N/2} } \binom{q_{N}}{2n}V^{2n} < (2V)^{q_{N}}e^{-\beta q_{N}}q_{N},
\end{eqnarray}
and the rational approximate pGFs converge for $V\lesssim e^{\beta}$. Thus, for $1<V<e^{\beta}$, both the horizontal and vertical unit cells are convergent -- this reproduces results in \cite{avila2017sharp}. 
\end{widetext}

\begin{figure*}
    \centering
    \includegraphics[scale = .54]{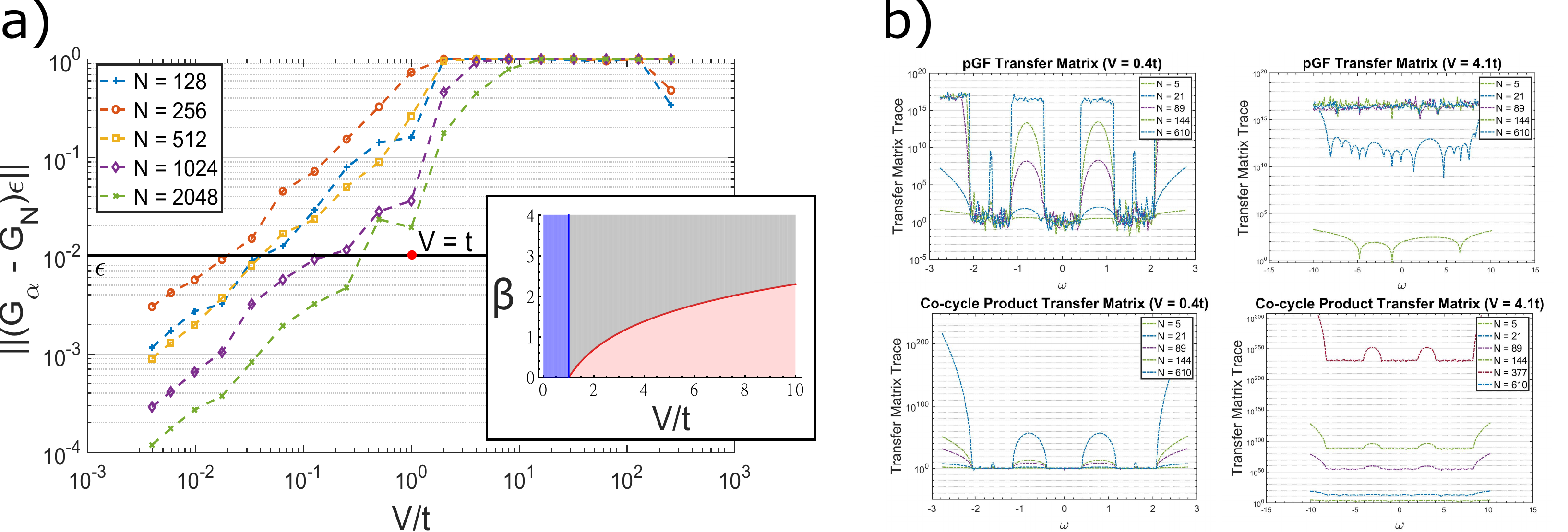}
    \caption{(a) Plot of max (operator norm) difference over all $\omega$ between the rational and irrational pGFs for $\alpha = (\sqrt{5}-1)/2$. Convergence is defined by the ability to take $\vert\vert G_{\alpha}(\omega+i\epsilon)-G_N(\omega+i\epsilon)\vert\vert <\epsilon^{2}.$ For $V/t<1$ ($V/t>1$) horizontal (vertical) unit cell converges. Taking larger system sizes N, sharpens convergence/divergence [shrinking divergence for large $V/t$ caused by finite system size]. (inset) AAH phase diagram for Liouville $\alpha$, with $\beta$ the Liouville exponent and transition at $\beta = \ln{(V)}$ (blue: metal, red: insulator, gray: transition).\\
    (b) Transfer matrix trace generated by pGF (top row) and direct calculation (bottom row) plotted for corresponding regimes, demonstrating convergence of Eq.~\eqref{tmeq}. Note the divergence of the trace for the insulating regime, implying only localized solutions.}
    \label{fig:convergence}
\end{figure*}

\subsection{S.I. Gauge Transformation}\label{gaugetransformsi}
If $V>1$, the above convergence proof fails and our rational Green's function approximates are no longer guaranteed converge to the irrational Green's function. Consequentially, the rational approximate transfer matrices no longer converge to the quasi-periodic transfer matrix and the quasi-periodic eigenfunctions are no longer the limit of the de-localized eigenfunctions of the rational approximate transfer matrices, defined by unit cells along the $x$-direction of the 2D parent Hamiltonian lattice. 

Instead we must apply a 2D gauge transformation to the 2D Hamiltonian, such that the ``magnetic" flux is contained in the $x$-direction and the unit cell is constructed in the $y$-direction. The resulting transfer matrix will be constructed from on-site projected Green's functions of the form
\begin{widetext}
\begin{eqnarray}
    G_{\perp,N} = \int\frac{d\phi}{2\pi}\left[\omega\mathbb{1}-\begin{pmatrix}
    2\cos{(2\pi \frac{p_{N}}{q_{N}}1+k_x)} & V &0& \ldots & Ve^{i\phi}\\
V & 2\cos{(2\pi \frac{p_{N}}{q_{N}}2+k_x)} & V& \ldots & 0\\
0 & V & 2\cos{(2\pi \frac{p_{N}}{q_{N}}3+k_x)}& \ddots& \vdots\\
\vdots & \vdots &\ddots& \ddots  & V\\
Ve^{-i\phi}  & \ldots  &0& V & 2\cos{(2\pi \frac{p_{N}}{q_{N}}q_{N}+k_x)}    \end{pmatrix} \right]^{-1}.\nonumber
\end{eqnarray}
We can factor out a V from the entire equation above and generate the same Green's function as above with $V\rightarrow 1/V$ and $\omega\rightarrow \omega/V$. As a consequence, we have a clear convergence for $1/V<1$ or $V >1$ for the projected Green's function along a vertical unit cell in the limit of $q_N\rightarrow\infty$.

As a subtle point, while this gauge transformation is well defined in 2D for Liouville numbers, it fails to manifest in 1D, as the ``magnetic flux" can only shift to the hopping elements in $\hat{H}$ under a quasi-Fourier transformation. When the rational approximates in the horizontal unit cell converge, the quasi-Fourier transform is close (exponentially for $\alpha$ Liouville) to a rational Fourier transform, and we simply obtain the horizontal unit cell rational approximates. This point only manifests itself for Liouville choices of $\alpha$, such that the convergence regime extends into the convergence regime of the gauge transformed Green's function, $V>1$.
\end{widetext}

For Liouville $\alpha$, multiple gauge choices and unit cell configurations define convergent approximations to the full quasi-periodic projected Green's functions, but the in the projection to 1D is biased towards the horizontal unit cells as the eigenfunctions survive this projection and the ``gauge group symmetry" -- phase shift invariance -- is preserved. 

More precisely, the Greshgorin circles force convergence for $V<1$, meaning any TME solutions must preserve the phase shift invariance. However, for Liouville irrationals, not only does a single unit cell converge, but any polynomial number of unit cells converge, i.e. $\delta_{N}q_{N}^{M} = e^{-\beta q_N}q_N^{M} \rightarrow 0$ in Eq.~\ref{diffbound}. Thus, even for the vertical unit cell choice, we preserve the $k_y$ quantum number in the 1D projection and we retain phase-shift invariance. Correspondingly, the phase diagram is biased towards the metallic phase in Fig.~\ref{fig:convergence}

\section{S.I. Projected Green's Function Formalism}\label{zerospolessi}
\indent In translation-invariant systems, the Brillouin zone allows for flexibility in writing locally computable formulas for topological invariants. In this language, Green's function zeros are singular and carry topological significance \cite{bernevig2013topological,Slager2015, Rhim2018, Borgnia2020, volovik2003universe, slager2019translational, Gurarie2011,mong2011edge}. More recently, it was noticed that bound state formation criteria along an edge are also defined by Green's function zeros \cite{Slager2015,Rhim2018,Borgnia2020,jitomirskaya2019critical,mong2011edge,volovik2003universe}, thereby tracking both topological invariants and their corresponding edge modes.

Extending this methodology beyond translation invariant systems consists of two steps. One must show both that Green's function zeros are still of topological significance and that edge formation criteria are still described by the presence of in-gap zeros. We first show the latter.

\indent The poles of the Green's function restricted to a particular site in position space correspond to an energy state at that particular site. Here restricted refers to the projection of the system Green's function, $G$, to a single site,
\begin{eqnarray}\label{gprojectsum}
G(\omega,\mathbf{r}_{\perp},\alpha_{\parallel}) = \sum_{\alpha} \vert\braket{\alpha\vert\mathbf{r}_{\perp}}\vert^{2}G(\omega,\alpha),
\end{eqnarray}
where $\alpha$ generically labels the Eigenvalues and $\alpha_\parallel$ is the remaining index post the contraction with $\mathbf{r}_{\perp}$. Generically, there will be many poles corresponding to the spectrum at $\mathbf{r}_{\perp}$, but they are not universal.

By adding on-site impurities and considering $G(\omega,\mathbf{r}_{\perp},\alpha_{\parallel})$ only in the band gap of the bare Green's function, any poles will be a result of the impurity potential, $\mathcal{V}(r) = \mathcal{V}\delta(r - \mathbf{r}_{\perp})$, binding a state in the gap. Then, by constructing an appropriate impurity geometry, and taking $|\mathcal{V}|\rightarrow\infty$, an edge is formed. Therefore, the condition for impurity localized states as $|\mathcal{V}|\rightarrow\infty$ is equivalent to the criteria for the formation of edge localized modes. And, impurity bound states correspond to zeros of the restricted in-gap Green's function. This is most readily seen by factoring the full Green's function, $G$ of some system with Hamiltonian $H_{0}$ and an impurity potential $\mathcal{V}$. That is, the full Green's function $G$ can be written in terms of the Green's function $G_{0} = (\omega - H_{0})^{\text{-}1}$ of the original system without the impurity, 
\begin{eqnarray}\label{factoring}
G(\omega,\alpha) &=& (\omega - (H_{0}+\mathcal{V}))^{\text{-}1}
=  (1+\mathcal{V}G_{0})^{\text{-}1}G_{0}.
\end{eqnarray} 
Correspondingly, impurity bound states (poles of $G$) in the gap (not a pole of $G_{0}$) must be a pole of $(1-\mathcal{V}G_{0})^{\text{-}1}$,
\begin{equation}\label{pgfdeteq}	\det \left[ G_{0}(\omega,\alpha)  \mathcal{V} - \mathbf{1} \right] = 0.
\end{equation}
For $|\mathcal{V}|\rightarrow\infty$, solutions require $G_{0} \rightarrow 0$, and the zeros of $G_{0}$ correspond to poles of $G$. Hence, the zeros of the restricted in-gap Green's function, $G(\omega,\mathbf{r}_{\perp},\alpha_{\parallel})$, correspond to edge modes, just as in the translation-invariant case \cite{Slager2015}.

\indent The above requires in-gap bound states of an aperiodic system to appear as zeros of the projected Green's function. We now derive the conditions under which such states are fixed by the pattern topology. In these cases, in-gap states survive small disorder \cite{prodan2015}, and impose constraints on system dynamics. The fundamental difference between the translation invariant and aperiodic cases comes down to the existence of a good momentum quantum number. For translation invariant systems, the $x$-basis is dual to the Brillouin zone momentum basis. Thus, each momentum eigenstate is equally weighted in the projection, Eq.~\eqref{gprojectsum} with position $0$, and singular points of the projected Green's function directly relate to an obstruction of consistently writing the Green's function over $k$-space. Heuristically, for two bands separated by a topologically non-trivial gap, the eigenstates switch eigenvalues \cite{Slager2015,Borgnia2020,Rhim2018}. 

There is no such guarantee for generic aperiodic systems, but we can reduce the the constraints of translation invariance to a single condition for which the sum in Eq.~\eqref{gprojectsum} is in fact reduce-able to a sum over the topologically-fixed IDoS. We now focus on 1D systems (higher dimensions generalize by choosing codimension-1 surface \cite{Borgnia2020,Slager2015,Rhim2018}), where the projected Green's function on the site $x_{0}$ reduces to 
\begin{eqnarray}\label{1Dgprojectsum}
G_{\perp}(\omega,x_{0}) = \bra{x_{0}}\left[\sum_{\alpha}G(\omega,\alpha)\ket{\alpha}\bra{\alpha}\right]\ket{x_{0}},
\end{eqnarray}
with $\alpha$ indexing the eigenstates of the system. In 1D, it is clear that any shift of the projection site $x_{0}$ can be absorbed as a phase shift in the pattern. We illustrate this with the AAH model, whose Hamiltonian reads
\begin{align}\label{aahrealspace}
	\hat{H} = \sum_{x} t(\hat{c}^{\dagger}_{x+1}\hat{c}_{x}+\hat{c}_{x+1}\hat{c}^{\dagger}_{x})+2V\cos(\Theta x+\delta_{x})\hat{c}^{\dagger}_{x}\hat{c}_{x}.
\end{align}
Here $\Theta$ is some irrational multiple of $2\pi$ generating the quasiperiodic pattern. The dynamics are generated by translations $\tau_{x}$, taking $V\cos(\Theta x+\delta_{x})\rightarrow V\cos(\Theta (x+1)+\delta_{x})$. And, we can shift $x_{0}\rightarrow x$ by taking $\delta_{x} = \Theta(x-x_{0})$. We can clearly index the Hamiltonian (and corresponding Green's function) by the real-space phase $\delta_{x}$. This is a general property of quasiperiodic patterns, and the spectrum is invariant under this shift \cite{bourne2018non}. This is guaranteed by choosing the hull of the pattern to define our unital algebra, as the system dynamics guarantee any initial point can be translated into any other point on the hull. We can therefore rewrite Eq.~\eqref{1Dgprojectsum} as
\begin{eqnarray}\label{1Dgprojectsumavg}
G_{\perp}(\omega,x_{0}) = \frac{1}{N}\sum_{x}\bra{x}\left[\sum_{\alpha}G(\omega,\alpha)\ket{\alpha(\delta_{x})}\bra{\alpha(\delta_{x})}\right]\ket{x},\nonumber\\
\end{eqnarray}
with $\delta_{x} = -\Theta(x-x_{0})$ depending on $x$, and the eigenstates depend on the choice of phase, i.e. an eigenstate localized at $x_{0}$ becomes localized at site $x$, but with the same energy and corresponding Green's function component, $G(\omega,\alpha)$. If this were a translation-invariant setting, it would be clear that the choice of $x_{0}$ cannot matter, and, thus the sum in Eq.~\eqref{1Dgprojectsumavg} must reduce to 
\begin{align}\label{1Dgprojectsum_nophase}
G_{\perp}(\omega,x_{0}) = \frac{1}{N}\sum_{\alpha}G(\omega,\alpha).
\end{align}
If this holds for the projected Green's function of an aperiodic system with non-trivial pattern topology, then the quantization of the IDoS fixes the sum in Eq.~\eqref{1Dgprojectsum_nophase}. All states are summed over. As a consequence, for an $\omega$ in a spectral gap, states above and below the gap contribute to the sum, such that for some $\omega_{*}$, $G(\omega)\vert_{\omega =\omega_{*}} = 0$. In translation invariant systems, the location of $\omega_{*}$ is usually protected by symmetries such as chirality, fixing states to be at equal energies above and below the gap. In quasiperiodic systems, the IDoS fixes the number of states above and below the gap\cite{bourne2018non}. For some gap labeled, $F$, the integrated density of states (IDoS) below each gap is fixed, i.e. for the AAH model $\text{IDoS}(F) = (m+n\Theta)\cup[0,N]$ \cite{bourne2018non}. Thus, for each gap, $F$, the relevant $\omega_{F}$ for which $G(\omega\in F)\vert_{\omega = \omega_{F}} = 0$ would also be fixed. This observation motivates our generalization to aperiodic systems. In particular, the ability to induce real-space translations as pattern phase shifts suggests that full translation invariance is not necessary.

\section{S.I. Projected Green's Function Zeros in 1D}
Having proven the pGF zeros are preserved in the irrational limit, we can project back down from 2D into 1D and construct the full solutions for the 1D problem. Notice this projection is natural with the horizontal unit cell choice and problematic in the gauge transformed vertical unit cells. 
\subsection{Horizontal Unit Cell}
For $V<t$ ($V< t e^{\beta(\alpha)}$ when $\alpha$ is Liouville), we proved the convergence of rational approximates to the incommensurate projected Green's function along a growing set of horizontal unit cells of size $q_{N}$, chosen via the continued fraction approximation for an irrational parameter, $\alpha$. In this context, a horizontal unit cell corresponds a periodic structure along the real-space coordinate direction, $x$ from the original 1D problem. As such the projection back down to 1D simply corresponds to the projection onto a single phase choice $\phi$ or $k_y$ in the rational approximates. Since phase is irrelevant in the limit of $N\rightarrow\infty$, the projected Green's function and by extension the transfer matrix of the 2D rational approximates is directly projected into 1D and the 2D irrational limit is equivalent to the 1D irrational transfer matrix.

Note, a direct consequence of this projection is the passing of bulk boundary correspondence via the zeros of the projected Green's function defining equivalence classes of transfer matrices and the existence of edge-localized modes as per Ref.~\cite{paper1}.

\subsection{Vertical Unit Cell}
For $V>t$ ($V>t e^{\beta(\alpha)}$), the convergent vertical unit cells are not as naturally projected into 1D. Choosing a particular phase, $\phi$, along the vertical unit cell corresponds to a momentum cut along $\delta_y$ in the 2D rational approximates. The cut is unnatural, however, because the $\delta_y$ quantum number doesn't survive the limit $N\rightarrow\infty$. Instead, we focus on the spectral gaps and the presence of pGF zeros.

As discussed above, pGF zeros survive the limit $N\rightarrow\infty$ and correspond to edge localized modes (rank deficient points in the transfer matrix). The rational approximates are known to host edge localized modes on open boundary conditions. The edge mode spectra span the bulk spectral gaps, filling the incommensurate spectral gaps in the limit $N\rightarrow\infty$. Thus, for each $\omega\notin\Sigma$ there exists a $\phi = \delta_y(\omega)$ such that 
\begin{align}
    G_{\perp,\alpha}(\omega,x_{0},\phi) = \lim_{N\rightarrow\infty} G_{\perp,N}(\omega,x_{0},\phi) = 0.
\end{align}

Projection to 1D now requires choosing a different $\delta_y$ for each site corresponding to the phase induced by lattice translation in the original quasi-periodic pattern. As all $\delta_y$ are included by the associated translations, if for any energy in the spectrum, $\omega \in \Sigma$, $\det{G_{\perp,N}(\omega,\delta_y)} = 0$, $\det{(\hat{T}_{q_{N}})} = 0$ and normalizable solutions do not exist (without an edge).

\subsection{Spectral Measure}
Bringing these ideas together, for any $\omega \in \Sigma$, there exists a $\gamma$ such that $\omega \pm \epsilon \notin \Sigma$ for any $0<\epsilon<\gamma$. In other words, the 1D spectrum for $V>t$ ($V>te^{\beta(\alpha)}$ is pure point-like as only the $\lambda$ such that $\det(\lambda\mathbb{1} -H_{\alpha})=0$ produce normalizable solutions, $\psi$, such that $\vert\vert \hat{T}\psi \vert\vert <\infty$, with $T$ the incommensurate transfer matrix. 

By contrast, for $V<t$ ($V< t e^{\beta(\alpha)}$, $\alpha$  Liouville) only one $\delta_y$ is fixed and not all $\omega$ are disallowed, one per gap in the limit of $N\rightarrow\infty$. More precisely, the 2D incommensurate transfer matrix is the same as the $T_{2D} = T_{1D}$. Thus, all 2D horizontal unit cell eigenfunctions are 1D eigenfunctions. And, since the translation invariant approximates converge in the limit of $N\rightarrow\infty$ the innerproduct measure of the 2D eigenstates is absolutely continuous by Floquet Theory \cite{avila2006reducibility,avila2006solving,avila2009ten,dinaburg1975one}. Thus, the 1D spectrum is abs. continuous for $V<t$ ($V<t e^{\beta(\alpha)}$).

Finally, the transition point ($V = t$ when $\alpha$ is Diophantine) is more complicated, and we save discussions for a future work. We do note that neither the horizontal nor vertical unit cells will converge in this case. We instead need a chiral gauge choice \cite{jitomirskaya2019critical}.

\subsection{S.I. AAH Algebra}\label{siaahalgebra}
\begin{widetext}
We follow the work of Prodan \cite{prodan2015} in deriving explicit topological invariants in the AAH model context. We construct the unital algebra and use it to label the resulting spectral gaps of the AAH Hamiltonian. Recall that it reads
\begin{eqnarray}\label{1Dham}
\mathcal{H}_{\delta_{x}} =\sum_{n}t \hat{c}_{n+1}^{\dagger}\hat{c}_{n} + \textnormal{h.c.} + 2V\cos(\Theta n+\delta_{x})\hat{c}_{n}^{\dagger}\hat{c}_{n},
\end{eqnarray}
in terms of the creation operators $c^{\dagger}_i$, lattice constant $a$, potential $V$ that depends on the position $n$ and is indexed by phase $\delta_{x}$.
The model exhibits a duality under the pseudo-Fourier transformation $c_{k} = \sum_{n} \exp(-ikn)c_{k}$  \cite{aubry1980annals}. Considering, $\delta_{x} = 0$, one obtains 
\begin{eqnarray}
    \tilde{\mathcal{H}}(k) &=& \sum_{k,k',n}t e^{ik(n+1)-ik'n}\hat{c}_{k}^{\dagger}\hat{c}_{k'}+t^{*} e^{ikn-ik'(n+1)}\hat{c}_{k}^{\dagger}\hat{c}_{k'} + V(e^{2\pi i an +i(k-k')n}+e^{-2\pi i an+i(k-k')n})\hat{c}_{k}^{\dagger}\hat{c}_{k'}\nonumber\\
  \tilde{\mathcal{H}}_{\phi}  &=& \sum_{k} 2t\cos(\Theta k)\hat{c}_{k}^{\dagger}\hat{c}_{k} + V(\hat{c}_{k+1}^{\dagger}\hat{c}_{k} +\textnormal{h.c.}).
\end{eqnarray}
where in the last line we have set $k = \Theta m$ and defined $\sum_{n} \exp(i\Theta n(m-m')) = \delta(m-m')$ in the limit $n\rightarrow\infty$.  
A natural equivalence emerges between $\mathcal{H}$ and $\tilde{\mathcal{H}}$ under $V\rightarrow t$, implying the model undergoes a transition for $V = t$, being the well known 1D metal-insulator transition. 
Considering the limits $V = 0$ and $t = 0$, the duality relates extended (momentum-localized) eigenstates to position-localized states. 

\indent The duality in the AAH model has been focus of many localization studies, past and present \cite{aubry1980annals,jitomirskaya2019critical}. The model took on new light, however, when \cite{kraus2012topological} noticed it could be parameterized by the phase choice $\delta_{x}$ \cite{kraus2012topological,jitomirskaya2019critical,avila2006solving,bellissard1982quasiperiodic}. Naively, this phase choice is irrelevant as it corresponds to a shift in initial position of an infinite chain, but the 2D {\it parent} Hamiltonian, as function of $x$ and  $\delta_{x}$, has a topological notion. In particular, it corresponds to a 2D tight-binding model with an irrational magnetic flux per plaquette. Explicitly,
\begin{eqnarray}\label{2Dhamiltonian2}
    \mathcal{H} &=&\sum_{n,\delta_{x}}t \hat{c}_{n+1,\delta_{x}}^{\dagger}\hat{c}_{n,\delta_{x}} + t^{*} \hat{c}_{n,\delta_{x}}^{\dagger}\hat{c}_{n+1,\delta_{x}} + 2V\cos(\Theta x+\delta_{x})\hat{c}_{n,\delta_{x}}^{\dagger}\hat{c}_{n,\delta_{x}},\nonumber\\
   \tilde{\mathcal{H}}  &=&\sum_{n,m,m'}t \delta_{m,m'}(\hat{c}_{n+1,m}^{\dagger}\hat{c}_{n,m'} + t^{*} \hat{c}_{n,m}^{\dagger}\hat{c}_{n+1,m'} )+V\left(e^{i\Theta x}\delta_{m+1,m'}+e^{-i\Theta x}\delta_{m-1,m'}\right)\hat{c}_{n,m}^{\dagger}\hat{c}_{n,m'},\nonumber\\
    \tilde{\mathcal{H}}  &=&\sum_{n,m}t (\hat{c}_{n+1,m}^{\dagger}\hat{c}_{n,m} + \hat{c}_{n,m}^{\dagger}\hat{c}_{n+1,m} )+ V(e^{i\Theta n}\hat{c}_{n,m+1}^{\dagger}\hat{c}_{n,m} + e^{-i\Theta n}\hat{c}_{n,m-1}^{\dagger}\hat{c}_{n,m}).
\end{eqnarray}
The 2D spectrum amounts to a Hofstadter butterfly when varying the flux per plaquette, $\Theta$. For any rational flux, $\Theta/2\pi = p/q \in\mathbb{Q}$, one can define a magnetic unit cell specifying bands that have a Chern number, which sum to zero. This is however not possible for an irrational flux. In this case strategies outlining sequences of rational approximates, with similar band gaps, to find topological invariants were employed \cite{PhysRevB.91.014108}. 

\end{widetext}
\indent Hamiltonian \eqref{2Dhamiltonian2} is manifestly topological for all rational fluxes, $a$ \cite{PhysRevB.91.014108}. We can therefore create a sequence of rational approximates to an irrational flux $\lbrace a_{n}\rbrace$, such that $\lim_{n\rightarrow\infty} a_{n} = a_{*}$. The problem, however, arises when projecting back down into one dimension. This is made most clear by considering the spectrum for different phase choices $\delta_{x}$, contrasting the irrational and rational case. For example, for $a = 1/2$, the choice $\delta_{x}$ changes the maximum amplitude of the on-site potential. In the irrational case, however, it has no effect on the spectrum and acts as a translation. Hence, the projection of the sequence of rational approximates {\it does not} create a 1D sequence of rational approximates to the AAH model \cite{prodan2015}. Instead more recent works leverage powerful tools from non-commutative geometry to tackle the problem conclusively, finding deep connections between non-commutative topological invariants and the inherited topology for the 1D projection, i.e the AAH model. In fact, methods described below describe both the inheritance of a 2D topological invariant and a bulk-boundary correspondence in the model \cite{prodan2015}. 

\begin{figure*}[ht]
    \centering
    \includegraphics[scale =.10]{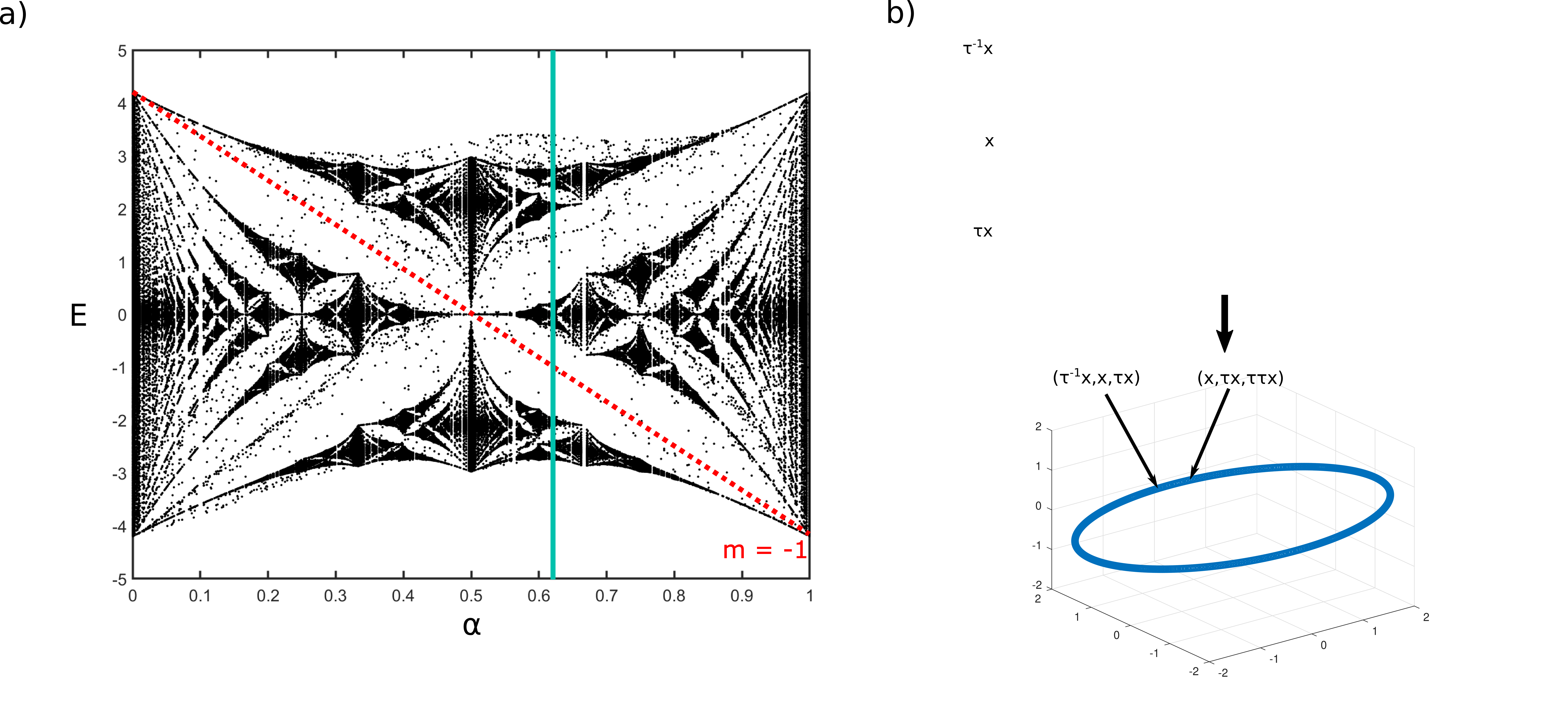}
    \caption{(a) Quasi-periodic spectrum as function of plaquette flux. Note the gap labeling, where gaps are generated by lines of different slopes (lines cured in spectrum, but straight in IDoS) (b) Illustration of quasi-periodic pattern generating a minimal surface (hull). This forms the underlying unital algebra, taking the place of a Brillouin zone.}
    \label{patternfigure}
\end{figure*}

\subsubsection{Non-commutative topological characterization}  We formalize the ideas behind a parent Hamiltonian for the AAH model and show the AAH model obeys the same algebra as the Hofstader Hamiltonian (defined below), forming a non-commutative torus. We show these two Hamiltonians are equivalent up to representation and all topological properties carry over \cite{prodan2015}.

\indent Consider the translation operator, acting on $\mathcal{H}_{\delta_{x}}$ as
\begin{align}\label{xspacetranslation}
    T^{n}\mathcal{H}_{\delta_{x}}T^{\dagger n} = \mathcal{H}_{\delta_{x}+n\Theta}.
\end{align}
For $\Theta/2\pi\notin\mathbb{Q}$, the repeated action of the translation operator parameterizes the motion of the  phase, $\phi\in\mathbf{S}$, along the unit circle. On the set of continuous functions over $\mathbf{S}$, $C(\mathbf{S}): \mathbf{S}\rightarrow\mathbb{C}$, the translation operator acts similarly. That is, for some $f \in C(\mathbf{S})$,
\begin{align}
    \alpha_{n}:f(\phi) \rightarrow f(\phi+\Theta n).
\end{align}
More formally, one paramterizes the action of $\mathbb{Z}$ on $\mathbf{S}$ as a dynamical system and then constructs its dual, i.e. the pattern hull described in the main text. The key step, however, is to define a unitary which acts in place of the translation operator on functions $f\in C(\mathbf{S})$.
\begin{align}
    u^{n}f(\phi)u^{-n} = f(\phi+\Theta n)
\end{align}
In this language, we can define elements of the space $C(\mathbf{S})\rtimes_{\alpha}\mathbb{Z}$, the semi-direct product between complex continuous functions of $\mathbf{S}$ and the translations $\mathbb{Z}$ generated by $\alpha: f(\phi) \xrightarrow{\alpha_{n}} f(\phi+\Theta n)$ as
\begin{align}
    \mathbf{a} = \sum_{n\in\mathbb{Z}} a_{n}u^{n}.
\end{align}
In the above $a_{n} \in C(\mathbf{S})$, and $u^{n}$ corresponds to a translation along $\mathbb{Z}$. The main benefit of defining this operator algebra corresponding to the 1D translations is that we can pick a representation of the Hilbert space,
\begin{align}\label{aahrep}
    \pi_{x}(\mathbf{a}) = \sum_{n,x\in\mathbb{Z}}a_{n}(\delta_{x}+\Theta x)\ket{x}\bra{x}T^{n},
\end{align}
with $\phi \in [0,2\pi]$. In this representation, the elements, $\mathcal{H}_{\delta_{x}}$, are simply represented as $\pi_{\phi}(\mathbf{h})$. Here, 
\begin{align}\label{Celement}
    \mathbf{h} = u+u^{-1} + 2V cos(\phi)
\end{align}
is an element of $C(\mathbf{S})\rtimes_{\alpha}\mathbb{Z}$, and the potential shifts when $T$ acts on $\phi$ at each site, i.e. $(\phi + \Theta n)\textnormal{mod}_{2\pi}$.

\indent As done for Eq. \eqref{2Dhamiltonian2}, we now show that this element $h\in C(\mathbf{S})\rtimes_{\alpha}\mathbb{Z}$ also generates the Hofstader Hamiltonian, see also \cite{prodan2015}. We first rewrite the Hofstader Hamiltonian as
\begin{align}\label{hofstaderhamiltonian2d}
    H_{\Theta} = \sum_{x,y} T_{x}+T_{x}^{-1}+V(T_{y}+T_{y}^{-1}),
\end{align}
where 
\begin{eqnarray}
T_{x}\ket{x,y} = \ket{x+1,y} \ \ \textnormal{and} \ \ 
T_{y} \ket{x,y} = e^{-i\Theta x}\ket{x,y+1}\nonumber
\end{eqnarray}
are magnetic translations with commutation relations $T_{x}T_{y} = e^{i\Theta}T_{y}T_{x}$. In this form, we define unitary operators $u$, as before, and $z = \exp{i\phi}$ acting on $C(\mathbf{S})$ corresponding to the translations along $x$ and $y$, respectively. These have the same commutation relations $uz = e^{i\Theta}zu$ and allow for a representation of the Hilbert space of $l^{2}$-normed functions on $\mathbb{Z}^{2}:$
\begin{align}\label{2delements}
    \pi' (\textbf{a}) = \sum_{n,m} f_{n}T_{x}^{n}T_{y}^{m}.
\end{align}
Then, $H_{\Theta} = \pi'(\textbf{h})$, for 
\begin{align}
    \mathbf{h} = u+u^{-1} + V(z+ z^{-1}) = u+u^{-1} + V (e^{i\phi} + e^{-i\phi})
\end{align}
as in Eq. \eqref{Celement}. 
\indent Framing the problem in terms of the operator algebra $C(\mathbf{S})\rtimes_{\alpha}\mathbb{Z}$ allows one to use techniques from non-commutative geometry to solve the problem immediately. In particular, this is a unital *-algebra for which a non-commutative calculus can be defined. The elements of the algebra are
\begin{align}
    \mathbf{a}  = \sum_{m,n\in\mathbb{Z}}f_{m,n}z^{m}u^{n},
\end{align}
\indent We can further revert to tools from non-commutative geometry to compute topological invariants. Details can be found in \cite{prodan2015}. We simply state the results here.
One can define differentiation intuitively along each direction:
\begin{eqnarray}
\partial_{1}\mathbf{a} &=& i\sum_{m,n\in\mathbb{Z}}m f_{m,n}z^{m}u^{n},\nonumber\\
\partial_{2}\mathbf{a} &=& i\sum_{m,n\in\mathbb{Z}}n f_{m,n}z^{m}u^{n}.
\end{eqnarray}
And, then integration follows as the inverse operation, $\mathcal{I}(\mathbf{a}) = f_{00}$, i.e. the constant term. These operations along with the algebra define the non-commutative Brillouin torus \cite{bellissard1986gaplabeling,bellissard1986k,bellissard2000hull}, $(C(\mathbf{S})\rtimes_{\alpha}\mathbb{Z},\partial,\mathcal{I})$, and form a special case of a spectral triple.

\indent Thus far, this section has only been a formalization of the concepts explained above. However, expressing the system as a spectral triple allows us to bring down the hammer of non-commutative geometry. In particular, there has been a careful formulation of K-theory in the case of spectral triples \cite{bellissard1986k,bourne2018non}. For the particularly simple case of a non-commutative torus, one can write down a locally computable index formula \cite{bourne2018non,bellissard1982quasiperiodic,bellissard1986gaplabeling,bellissard2000hull,bellissard1986k}, and compute a Chern number.
We introduce a projection operator, $\mathbf{p} = 1/2(1+\sgn(\epsilon_{F}-\mathbf{h}))$, which defines a filling of the spectrum below some Fermi level, $\epsilon_{F}$. The first non-commutative Chern number is then given by\cite{prodan2015,bellissard1986gaplabeling,bellissard1986k,bellissard2000hull,prodan2013non}
\begin{align}
    \textnormal{Ch}_{1}(\mathbf{p}) = 2\pi \mathcal{I}(\mathbf{p}[\partial_{1}\mathbf{p},\partial_{2}\mathbf{p}]).
\end{align}
Like the normal Chern invariant, this is well defined as long as there is a finite spectral gap.
Here we replicate the key result of \cite{prodan2015}, by computing this integral in the representation given by Eq. \eqref{aahrep}.
where $\mathcal{I}(\mathbf{a}) = 1/(2\pi)\int_{\mathbf{S}}d\phi f_{0}(\phi)$. In this simple case, the integral reduces to 
\begin{align}
    \mathcal{I}(\mathbf{a}) &=& \lim_{N\rightarrow\infty} \frac{1}{2N}\sum_{-N\leq x\leq N}f_{0}(\phi+\Theta x) =  \tr_{L}(\pi_{\phi}(\mathbf{a}))
\end{align}
where we have used that $f_{0}(\phi+\Theta x) = \bra{x}\pi_{\phi}(\mathbf{a})\ket{x}$, and $\tr_{L}$ is the normalized trace. In this representation
\begin{align}
    \textnormal{Ch}_{1} = 2\pi i \tr_{L}\left(\pi_{\phi}(\mathbf{p}[\partial_{1}\mathbf{p},\partial_{2}\mathbf{p})\right)
\end{align}
Using that, 
\begin{eqnarray}
\pi_{\phi} (\partial_{1}\mathbf{p})=\partial_{\phi}\pi_{\phi}(\mathbf{p}) \ \ \textnormal{and} \ \ \pi_{\phi} (\partial_{2}\mathbf{p})=i\left[X,\pi_{\phi}(\mathbf{p})\right],\nonumber
\end{eqnarray}

Defining $P_{\phi} = \pi_{\phi}(\mathbf{p})$, the projection operator in the AAH representation, the Chern number takes on a simple form,
\begin{align}
    \textnormal{Ch}_{1} = -2\pi  \tr_{L}\left(P_{\phi}[\partial_{\phi}P_{\phi},[X,P_{\phi}]]\right).
\end{align}
\indent Therefore, the Hofstader and AAH Hamiltonians are generated by the same element of $C(\mathbf{S})\rtimes_{\alpha}\mathbb{Z}$. This proves that the topological invariant of the 2D Hofstadter Hamiltonian is inherited by the 1D AAH model and explains the natural appearance of bulk-boundary correspondence -- the existence of boundary localized states reflecting the bulk topological invariant. The topological invariant is robust to disorder, and the edge spectrum is gapless when cycling through $\phi$ \cite{prodan2015}. 

An interesting consequence of this pGF transfer matrix formalism is the clear connection via the transfer matrix poles between the non-commutative geometry of the system and the spectral measure. Interestingly, this topological criteria was also noticed by \cite{jitomirskaya2019critical} using a different set of techniques to analyze the semi-infinite AAH model.

\subsection{S.I. Gap-Labeling Theorems}\label{gltheoremsi}
A well studied question arises from this topological criterion. Non-commutative geometry predicts the existence of gaps in the quasi-periodic integrated density of states (IDoS) depending on the irrational parameter $\alpha$ in the AAH Hamiltonian. Given the clear role topology plays in the dynamics, it was predicted and shown that these gaps in the IDoS form open sets in the complement of the quasi-periodic spectrum \cite{bellissard1986gaplabeling,bellissard1982quasiperiodic,bellissard1986k}.

Quasi-periodic systems can be indexed by patterns generating a "deterministic" disorder \cite{bourne2018non,bellissard2000hull,prodan2013non}. In Fig.~\ref{patternfigure}b peaks can be labeled by a coordinate, $P = \lbrace p_{i}\rbrace_{i\in\mathbb{Z}}$, forming a pattern. In the absence of a Brillouin zone, we consider the pattern as a dynamical system and find its convex hull - the minimal surface into which it can be embedded, $\Omega$. For example, for a simple generator such as $G = \cos(\Theta x)$ with $\Theta/2\pi \notin\mathbb{Q}$ it forms a ellipse. However, we can similarly find the hull for more complicated patterns upon defining the map, $f = \lbrace p_{i} \in P\vert f(p_{i}) = (p_{i+1} - p_{i},p_{i+2}-p_{i+1},\ldots)\in X\rbrace$ where $X$ is a hyper-cube of edge length defined by the pattern $P$. Although each element of the pattern is assigned a coordinate in an arbitrarily high dimensional space, there are only as many linearly independent coordinates as there are generators of the pattern. For the aforementioned sinusoidal generator of fixed amplitude, the element $p_{i+1} - p_{i}$ sets the period and further elements - $p_{i+2}-p_{i+1},\ldots$ - are linearly dependent on the first two by a translation. Consequently, it forms the anticipated ellipse in any dimensional hypercube rather than a higher dimensional surface, see Fig.~\ref{patternfigure}b. Ergodicity on this minimal embedding implies there exists a trajectory between any initial approaching (arbitrarily close) any other point on the surface. On the pattern hull, the notion of gauge invariance for quasiperiodic eigenstates mentioned above is similar to gauge invariance in a Brillouin zone, taking $\ket{k}\rightarrow\ket{k+\delta}$.

\indent Thus, we consider the space of continuous functions on the hull of the pattern $\mathcal{C}(\Omega)$, the direct analog of a Brillouin zone, and introduce dynamics by defining the action of pattern translations, $\tau$, on these functions, defining a so-called $C^{*}$-Algebra, $\mathcal{C}(\Omega)\rtimes_{\tau}\mathbb{G}$, where $\mathbb{G}$ is the group generated by translations. Elements of this unital algebra are the non-commutative analogs for Hamiltonians on the momentum space torus. Adding information about the on-site Hilbert space - bands in translation invariant case - and a differential operator extracts sufficient information from the quasiperiodic system to define topology in the same way as done in conventional translation invariant band topology. More specifically, one relies on a generalization of the Atiyah-Singer Index theorem to spectral triples by Connes and Teleman \cite{connes1994quasiconformal}. 

\indent As discussed in S.I.\ref{siaahalgebra}, the unital algebra generated by a quasiperiodic pattern with real-space translations is a non-commutative n-torus -- dimension corresponding to the number of generators for the pattern and system dynamics. As a direct consequence, the spectral gaps of quasiperiodic systems as a function of the incommensurate parameter, $\Theta \in [0,1]$, can be labeled by integers, see Fig.~\ref{patternfigure}a, i.e. $\lbrace m+n\Theta\vert m,n\in\mathbb{Z}\rbrace\cap[0,N]$ with N is system size, label the gaps in the IDoS of the AAH Model. By construction, this labeling is invariant to small disorder as long as the pattern is well defined and gaps remain open \cite{bourne2018non,prodan2015,prodan2013non}.

\end{document}